\documentclass[prb,amsmath,amssymb,reprint,longbibliography,float,floatfix]{revtex4-1}
\usepackage{graphicx}
\usepackage{dcolumn}
\usepackage{color}
\usepackage{soul}
\usepackage{float}
\usepackage{times}
\usepackage{xspace}
\usepackage[pdftex,breaklinks,colorlinks,linkcolor=blue,anchorcolor=blue,citecolor=blue,urlcolor=blue]{hyperref}
\usepackage{lscape}
\usepackage{nameref}
\setcitestyle{numbers,round,super}

\newcommand{\del}[1]{\!\xspace}

\usepackage[fulladjust]{marginnote}
\newcommand{\mpar}[1]{}

\newcommand{\mnote}[1]{}

\newcommand{\new}[1]{{\color{black} #1}}

\usepackage[columnwise]{lineno}
\nolinenumbers

\def\k{{\bf k}}

\def\d{\mathrm{d}}
\def\x{{\bf x}}

\makeatletter 
\renewcommand{\eqref}[1]{\textup{{equation~(\ref{#1})}}}
\newcommand{\tabref}[1]{\textup{{Table~\ref{#1}}}}
\newcommand{\figref}[1]{\textup{{Fig.~\ref{#1}}}}

\makeatother

\makeatletter
\def\ps@pprintTitle{%
   \let\@oddhead\@empty
   \let\@evenhead\@empty
   \def\@oddfoot{\reset@font\hfil\thepage\hfil}
   \let\@evenfoot\@oddfoot
}
\makeatother

\makeatletter
\AtBeginDocument{\let\LS@rot\@undefined}
\makeatother

\usepackage{etoolbox}
\makeatletter
\renewcommand\@biblabel[1]{\hspace*{\labelwidth}}
\apptocmd{\NAT@thebibliography}{\setlength\itemindent{-6pt}}{}{}
\makeatother
\begin{document}

\title{A time-dependent diffusion MRI signature of axon caliber variations and beading}

\author{Hong-Hsi Lee}
\email{Honghsi.Lee@nyulangone.org}
\author{Antonios Papaioannou}
\author{Sung-Lyoung Kim}
\author{Dmitry S. Novikov}
\author{Els Fieremans}
\address{Center for Biomedical Imaging and Center for Advanced Imaging Innovation and Research (CAI$^2$R), Department of Radiology, New York University School of Medicine, New York, NY 10016, USA}

\begin{abstract}
\noindent
MRI provides a unique non-invasive window into the brain, yet is limited to millimeter resolution, orders of magnitude coarser than cell dimensions. Here we show that diffusion MRI is sensitive to the micrometer-scale variations in axon caliber or pathological beading, by identifying a signature power-law diffusion time-dependence of the along-fiber diffusion coefficient. We observe this signature in human brain white matter, and uncover its origins by Monte Carlo simulations in realistic substrates from 3{\it d} electron microscopy of mouse corpus callosum.  Simulations reveal that the time-dependence originates from axon caliber variation, rather than from mitochondria or axonal undulations.  We report a decreased amplitude of time-dependence in multiple sclerosis lesions, illustrating the sensitivity of our method to axonal beading in a plethora of neurodegenerative disorders.  This specificity to microstructure offers an exciting possibility of bridging across scales to image cellular-level pathology with a clinically feasible MRI technique.

\end{abstract}

\date{\today}


\maketitle

\section*{Introduction}
Diffusion MRI (dMRI) is sensitive to the micrometer length scale via the commensurate diffusion length, and as such, is a promising in vivo technique for evaluating micrometer-scale structural features (the so-called tissue microstructure) of biological tissues in health and disease. The sensitivity to tissue microstructure, however, is indirect, due to averaging of the local diffusion propagator over the millimeter-sized MRI imaging voxel. Biophysical modeling of the diffusion signal in biological tissue \citep{grebenkov2007review,jones2010book,kiselev2017mrphysics,novikov2019review} is therefore essential for quantification of cellular parameters, and to gain specificity to cellular changes in development, aging and pathology. This raises the critical question of which salient features of cells or tissues can be robustly retrieved across the gap of three orders of magnitude in spatial scales, and what the essential assumptions are to construct the most parsimonious biophysical models, thereby attaining the highest precision without losing accuracy.

Axonal microgeometry in brain white matter (WM) is special, as axonal diameters are much thinner than the clinically attainable diffusion length $L_d(t)\sim 10\,\mu$m. Hence, intra-axonal diffusion has been described \citep{kroenke2004} as occurring within infinitely narrow featureless impermeable tubes --- dubbed ``sticks" --- inside which diffusion is  effectively one-dimensional and Gaussian, completely determined by a constant diffusion coefficient. 
This simplified viewpoint --- a cornerstone ingredient of the so-called WM Standard Model \citep{novikov2019review} --- has been the basis for white matter dMRI modeling over more than a decade 
,   approximating the net intra-axonal space (IAS) within an MRI voxel as a collection of these sticks. In this picture, sticks are deemed non-exchanging with extra-axonal water, and their overall orientation is modeled either by a specific distribution function, such as the Watson distribution \citep{zhang2012noddi}, or by using spherical harmonics 
\citep{jespersen2007sh,jespersen2010sh,reisert2017bayesian,novikov2018rotinv,veraart2019highb}. The stick model parameters, such as the the intra-stick diffusion coefficient and the orientation dispersion, provide biophysical significance, as they make dMRI specific to  axonal pathology. 

While suggested by NAA experiments 15 years ago \citep{kroenke2004}, for water dMRI the stick picture has been validated only recently.
Such validation is challenging, since fit quality alone is insufficient to validate a model. 
Selecting models becomes feasible by testing their unique functional forms in the domain where the dependence on experimental parameters clearly reveals their assumptions \citep{novikov2018onmodeling}. Borrowing this methodology from the physical sciences, the assumptions of the existence of sticks (i.e., of the locally $1d$ water diffusion), and of negligible exchange between sticks and extra-axonal water on the time scale of clinical dMRI, have  been validated in vivo in human brain WM by observing the $1/\sqrt{b}$ dMRI signal scaling (ideal stick response) at very strong diffusion weighting $b\sim 10\,\mathrm{ms/\mu m^2}$. \citep{mckinnon2017highb,veraart2019highb}

How adequate is the picture of featureless sticks? 
In this work, we show that the diffusion inside the IAS along axons is non-Gaussian at clinically employed diffusion times $t\sim 10-100\,$ms, and identify the dominant geometric features for this non-Gaussianity which can thus be quantified with a  dMRI measurement. 
For that, we focus on varying the diffusion time $t$, rather than on increasing the dMRI wave vector $q$. 

The absence of time-dependence in the overall diffusivity $D$ would signify Gaussian diffusion in every tissue compartment, while the presence of $t$-dependence would reveal  microscopic heterogeneity being coarse-grained by diffusion in at least one of the compartments \citep{novikov2010emt,novikov2014meso,novikov2019review,kiselev2017mrphysics}. So far, many WM studies focused on the diffusion time-dependence {\it perpendicular} to axons  to probe
the inner axon diameter \citep{assaf2008axcaliber,barazany2009axcaliber,alexander2010activeax,desantis2016extra} and the packing correlation length of the extra-axonal space \citep{burcaw2015meso,fieremans2016invivo,lee2018rd,desantis2016extra}.
Recently, however, the diffusion tensor eigenvalue {\it parallel} to major human WM tracts was found to decrease by 10-15\% over the range $t = 50-600\,$ms  using stimulated-echo dMRI \citep{fieremans2016invivo}. 
This non-trivial time-dependence along the tract could not be explained solely by the fiber dispersion, 
 i.e., by the locally transverse $t$-dependent contributions projected onto the tract direction. 
Rather, the observed non-Gaussian diffusion along axons suggests that either the extra-axonal space, or the IAS (the sticks) should be augmented to incorporate micrometer scale restrictions {\it along} the axon bundle direction. 

\begin{figure*}[th!!]
\centerline{\includegraphics[width=0.75\textwidth]{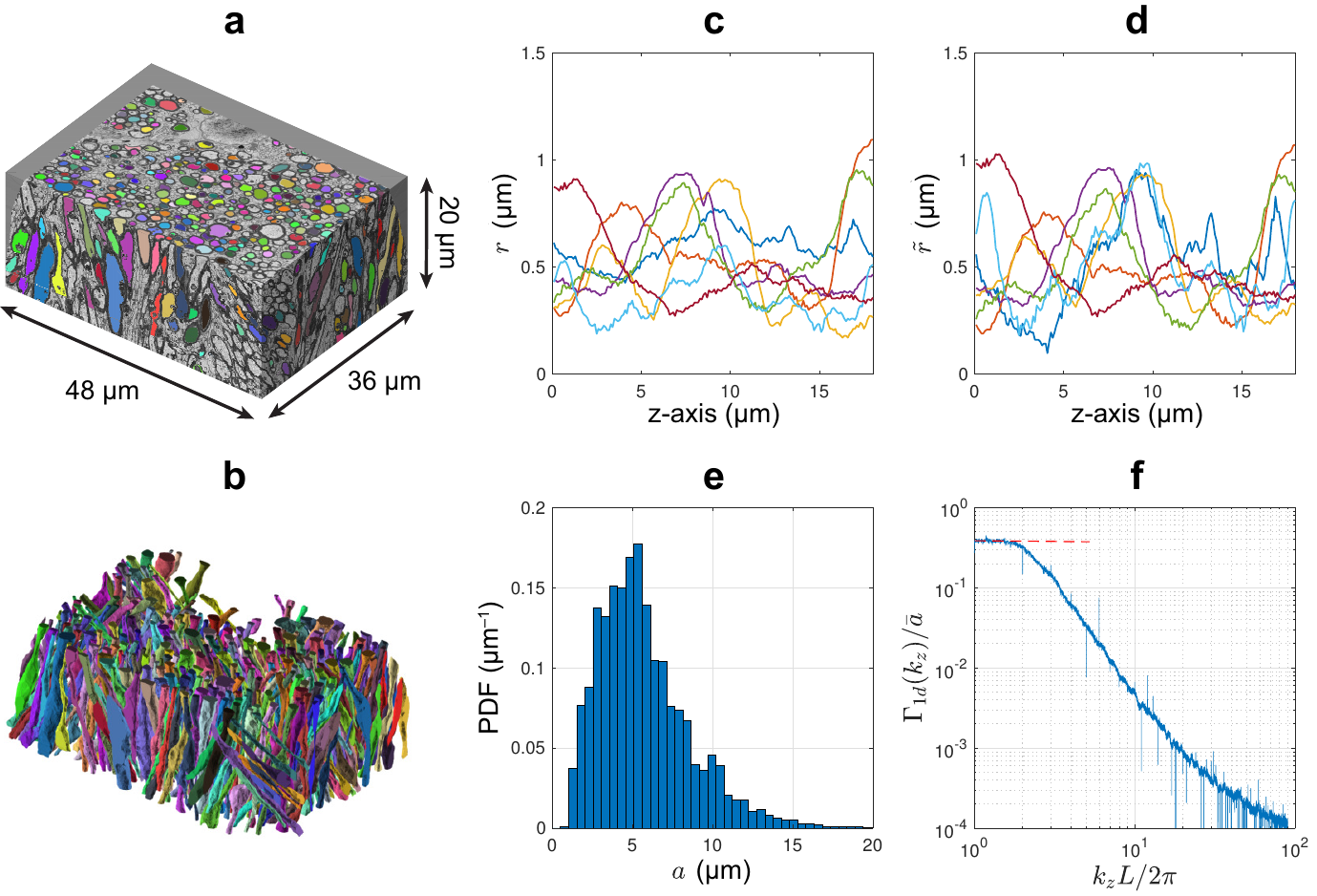}}
\caption{Structural analysis of axons segmented from female mouse brain corpus callosum EM reveals that the 1{\it d} placement of caliber variations exhibits {\it short-range disorder}, characterized by a finite correlation length. 
{\bf a} $3d$ EM image with segmented axons passing through the central slice. 
{\bf b} $3d$ representation of the intra-axonal space (IAS) segmentation yielding 227 axons that are long enough to pass through all slices ($>$ 20 $\mu$m). 
{\bf c} Radius variation $r(z)$ and 
{\bf d} normalized radius variation $\tilde{r}(z)$ along 7 selected axons. 
{\bf e} Histogram of distances $a$ between local radius maxima along all 227 segmented axons. 
{\bf f} The power spectrum $\Gamma_{1d}(k_z)$ along all axons shows a plateau at low $k_z \sim 1/L$ with $L$ the concatenated axon length (red dashed line), indicating a structural exponent 
$p = 0$, which corresponds to the short-range disorder, and leads to the dynamical exponent $\vartheta = 1/2$ in \eqref{eq:Dt-overall}, cf. \eqref{eq:universality}. 
Panels {\bf a} and {\bf b} are adapted from ref. \citep{lee2019em} by permission from Springer Nature: Springer Nature, Brain Structure and Function, Copyright, 2019.}
\label{fig:ias-shape}
\end{figure*}

What are these restrictions? 
According to the effective medium theory of ref. \cite{novikov2014meso}, 
observing a specific {\it power-law} time-dependence of the 
$1d$ diffusivity (along fiber)
\begin{equation} \label{eq:Dt-overall}
D(t)\simeq D_\infty + c\cdot t^{-\vartheta}\,, \quad \vartheta = \frac12
\end{equation}
approaching its long-time limit $D_\infty$ with the strength $c$ of restrictions, 
is a signature of a {\it short-range disorder} of the placement of the restrictions. However, this general theory does not reveal the exact source of these restrictions, be they the mictochondria, or beads, or axonal undulations, or the disordered extra-axonal space geometry.

Here we show that the IAS diffusion $t$-dependence has the form (\ref{eq:Dt-overall}), and is most sensitive to {\it axon caliber variations}, a vital signature of normal axonal microgeometry, that may be altered in pathology, turning into axonal beading. 
The  strength $c$ of restrictions to axial diffusion emerges from randomly-placed local axon caliber maxima
and depends on caliber variation.
To verify the power-law (\ref{eq:Dt-overall}) and attribute it to axon caliber variation,  we evaluate the effect of  axon shape on diffusion by developing, for the first time to our knowledge, full  3-dimensional ($3d$) Monte Carlo (MC) simulations of dMRI in a realistic microgeometry based on $3d$ electron microscopy (EM) segmentation \citep{lee2019em} of mouse brain corpus callosum. 

This paper is organized as follows. Firstly, we link the power-law dynamics of the time-dependent diffusivity along axons, $D(t)$, with the power spectrum of restrictions to $1d$ diffusion, allowing us to predict the time-dependence from the EM-derived axonal structure. Further, we perform MC simulations in realistic $3d$ IAS to calculate dMRI-related metrics, such as $D(t)$ and time-dependent kurtosis, $K(t)$, and study their unique functional forms. To better understand the origin of the diffusion time-dependence along axons, we separately evaluate the effect of mitochondria, caliber variation, undulation, and axonal orientation dispersion. Our simulations reveal axon caliber variation as the dominant source of time-dependent diffusion along axons. Finally, we show that theory and simulations are consistent with in vivo brain data in 15 healthy subjects acquired using pulsed-gradient spin-echo (PGSE) dMRI at clinically diffusion times. The change of diffusivity time-dependence due to specific pathology is also demonstrated in pilot data of 5 multiple sclerosis patients. In conclusion, we connect our theory, MC simulations, and clinical dMRI measurements into an overarching picture of a fundamental biophysical phenomenon ---  axonal caliber variation manifested by a signature power-law exponent $\vartheta = 1/2$ --- providing a remarkable specificity of a macroscopic dMRI measurement to a particular geometric feature of micrometer-scale axonal microstructure.

\cleardoublepage
\section*{Results}

\subsection*{From axonal structure to the diffusive dynamics} 
\label{sec:methods-power-spectrum}

The power-law tail of the diffusion time-dependence,  \eqref{eq:Dt-overall}, is determined by the {\it structural universality class} of the medium \citep{novikov2014meso}, with dynamical exponent 
\begin{equation} \label{eq:universality}
\vartheta =\frac{p+d}{2}
\end{equation}
in $d$ spatial dimensions. 
It was noted  that randomly looking media can be random in a few distinct ways, and thereby can be classified into a few so-called universality classes (analogously to the universality classes in the theory of critical phenomena). 
A structural universality class is defined by the structural exponent $p$, describing the statistics of long-range structural fluctuations. Technically, $p$ is defined via the asymptotic behavior 
\begin{equation*}
\Gamma(\k) = \int\! \mathrm{d}^d{\bf x}\, \Gamma({\bf x}) \, e^{-i{\bf k}{\bf x}} \sim k^p \,, \quad k\to 0
 \end{equation*}
 of the {\it power spectrum} $\Gamma(k)$ of the medium at low wave vector --- equivalently, the asymptotic behavior of 
 the density-density correlation function 
 \begin{equation} \label{Gamma-r}
 \Gamma(\x) = \langle \rho(\x_0+\x)\rho(\x)\rangle_{\x_0}
 \end{equation}
 at large distances $|\x|$ (here, the average $\langle ...\rangle$ is performed over the initial point ${\bf x}_0$). 
Molecular displacement over the diffusion length $L_d(t)$  probes the distances $|\x|\sim L_d(t)$, and thereby samples the statistics of spatial density fluctuations. Thus, \eqref{eq:universality} provides the fundamental connection between structure and dynamics. 
 
To determine the structural universality class of the microgeometry along axons, we begin from the $d=3$ 
density-density correlation function, \eqref{Gamma-r}, where $\rho({\bf x})$ is the $3d$ binary mask of an axially symmetric cylinder with radius variation $r(z)$ along axonal axis $z$. We would like to construct the corresponding $d=1$ power spectrum 
\begin{equation} \label{Gamma1d=Gamma3d} 
\Gamma_{1d}(k_z)=\frac{1}{\,\overline{A}\,} \int \Gamma_{3d}({\bf x})\, e^{-ik_z z} \d^3{\bf x} 
=\left.\frac{\left|\rho(\k_\perp, k_z)\right|^2}{V\cdot \overline{A}}\right|_{\k_\perp=0}
\end{equation} 
relevant at long distances $\sim 1/k_z$ exceeding the transverse dimensions of axons, when the diffusion becomes effectively one-dimensional. 
Hence, in \eqref{Gamma1d=Gamma3d}, $\k_\perp = (k_x, \ k_y)$ is set to $0$ as the diffusive motion is fully coarse-grained within the axonal cross-section on time scales much faster than the relevant diffusion times. In this Equation, we also used the Wiener-Khinchin theorem $\Gamma(\k) = |\rho(\k)|^2/V$, where $V$ is the (axonal) volume. Finally, since our resulting object $\Gamma_{1d}(k_z)$ is a $1d$ power spectrum, it should have dimensions of length, hence we normalize by the mean cross-sectional area $\overline{A}$.

The restrictions in general can be provided by any kind of microstructural inhomogeneity. 
Here, they are interpreted as coming from focal swellings or beads (caliber maxima)  and constrictions (minima) along axons. 
Below we study the behavior
\begin{equation} \label{eq:power-spectrum}
\Gamma_{1d}(k_z)|_{k_z\to0}\sim k_z^p\,, \quad k_z \to 0
\end{equation}
which will determine the  structural exponent $p$ determining the universality class of the $d=1$ microgeometry.

\subsection*{Axonal structure analysis reveals short-range disorder}

To estimate the structural exponent $p$ in \eqref{eq:power-spectrum}, 
we calculate the power spectrum using the radius variation along 227 segmented myelinated axons aligned with the $z$-axis (\figref{fig:ias-shape}a-b). \citep{lee2019em}
Practically, each axon's inner radius variation $r(z)$ (\figref{fig:ias-shape}c) is first normalized by the mean and the standard deviation of radii of all axon segments (standard score, \figref{fig:ias-shape}d). Next, the normalized radius variations are randomly concatenated along the $z$-axis. Finally, the concatenated normalized radius variation is rotated around the $z$-axis to generate an axially symmetric 3{\it d} binary mask $\rho({\bf r})$. 
The $1d$ power spectrum $\Gamma_{1d}(k_z)$ is calculated according to \eqref{Gamma1d=Gamma3d}. 

The power spectrum $\Gamma_{1d}(k_z)$ (\eqref{eq:power-spectrum}) along the concatenated axon with normalized radii  approaches a {\it plateau} at low $k_z$ (\figref{fig:ias-shape}f), and indicates a structural exponent 
\begin{equation} \label{p}
p = 0 \,.
\end{equation}
\eqref{eq:universality} thus yields dynamical exponent $\vartheta= 1/2$ in dimension $d=1$. 
Our {\it prediction} of $\vartheta=1/2$ and of the power-law tail in \eqref{eq:Dt-overall} will be tested below using MC simulations and dMRI measurements in human subjects.

The low-$k_z$ plateau demonstrates that restrictions along axons are randomly distributed with a finite correlation length, which is by definition a short-range disorder class of randomness. The level of the plateau is determined by the mean $\bar{a}$ and the variance $\sigma_a^2$ of the distance between restrictions (see Eq. [S13] and following derivations in the Supplementary Information of ref. \citep{novikov2014meso}), as well as by the average restriction width $\bar{l}$ (see Eq. (47) in ref. \citep{burcaw2015meso} with restriction ``shape" $v(k)|_{k\to0}$ $\to\bar{l}$):
\begin{equation} \label{eq:power-spectrum-plateau}
\Gamma_{1d}|_{k_z\to0,p=0}\simeq \frac{\sigma_a^2}{\bar{a}^2} \cdot \frac{\bar{l}^2}{\bar{a}}\,.
\end{equation}
The normalized power spectrum in \figref{fig:ias-shape}f has a low-$k_z$ plateau at $\Gamma_{1d}(k_z)/\bar{a}\approx$ 0.38, corresponding to an average restriction width $\bar{l}\approx$ 7.0 $\mu$m, for which $\bar{a}\approx$ 5.70 $\mu$m and $\sigma_a\approx$ 2.88 $\mu$m (\figref{fig:ias-shape}e) are estimated by locating the local maxima of axon caliber variations.

\begin{figure*}[th!!]
\includegraphics[width=1.0\textwidth]{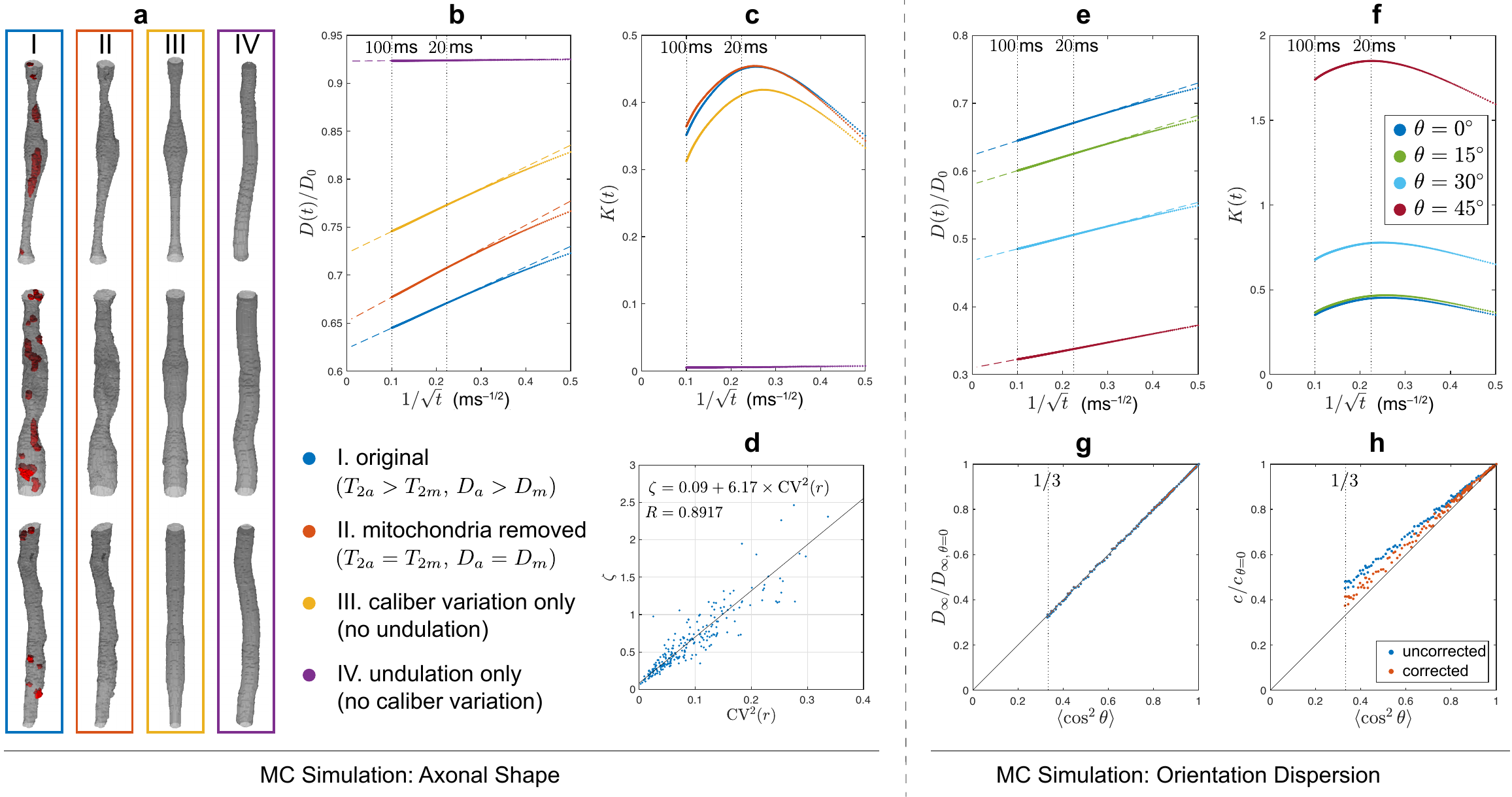}
\caption[]{\textit{MC Simulations inside 227 axons to study time-dependent diffusion along axons: \bf{Which microstructural axonal shape feature explains the observed diffusion $t$-dependence?}} {\bf a} Starting from the geometry of axons segmented from EM, four different types of microgeometries were created as follows (see text):  (I) with or (II) without considering shorter transverse relaxation time $T_{2m}$ and smaller intrinsic diffusivity $D_m$ in mitochondria (red), and derived synthetic axons with (III) only caliber variation or (IV) only axonal undulation.
{\bf b} The simulated $D(t)$ along axons for scenarios (I-IV) plotted as $1/\sqrt{t}$.  The linear scaling at long $t$ points to $1d$ short-range disorder along axons. Dashed lines are the asymptotes based on \eqref{eq:Dt-overall} and \eqref{eq:D-c-overall}, with fit parameters $D_\infty$ and $c$  in \tabref{tab:ias-simulation-shape}. 
Remarkably, $D(t)$ is mostly influenced by caliber variation, as it becomes much weaker in scenario IV when caliber variations are removed. 
{\bf c} The simulated $K(t)$ along axons for scenarios (I-IV) plotted as $1/\sqrt{t}$, showing a non-monotonic change with $t$. While removing undulations (III) slightly lowers the kurtosis, its significant reduction and altering the $t$-dependence occurs when  caliber variations are removed (IV). Both {\bf b} and {\bf c} indicate that axon caliber variations are the dominant contributions to the IAS time-dependence.  
{\bf d} Illustration of the relation $\zeta \propto \text{CV}^2(r)$ in the original IAS (scenario I), see \eqref{eq:app-cv} derived in {\it Methods} using coarse-graining arguments. 
\textbf{What is the effect of Orientation Dispersion?} {\bf e} Axon bundles with axially-symmetric orientation dispersion were created with orientation distributions of polar angles $\theta$ = [0$^\circ$, 15$^\circ$, 30$^\circ$, 45$^\circ$]. The simulated $D(t)$ along axons scales as $1/\sqrt{t}$ and decreases with the dispersion angle. The dashed lines are predictions based on \eqref{eq:Dt-overall} and \eqref{eq:D-c-overall}, with parameters $D_\infty$ and $c$ shown in \tabref{tab:ias-simulation-dispersion}. {\bf f} The simulated $K(t)$ along axons increases with the dispersion angle. {\bf g} The bulk diffusivity for $t\to\infty$, $D_\infty \propto\langle\cos^2\theta\rangle$, \eqref{eq:Dt-D-c-theta}. {\bf h} The strength of restrictions, $c$, slightly deviates from this proportionality relation in \eqref{eq:Dt-D-c-theta} (blue). Accounting for the higher-order $1/t$ term in \eqref{eq:Dt-modify}, the corrected value of $c$ restores \eqref{eq:Dt-D-c-theta} (red).
}
\label{fig:ias-simulation-shape}
\end{figure*}


\subsection*{Simulations validate time-dependent diffusion due to caliber variation }

\begin{table}[th!]
\centering
\begin{tabular}{l|c c}
\hline
Microgeometry (\figref{fig:ias-simulation-shape}a) & $D_\infty$ ($\mu$m$^2$/ms) & $c$ ($\mu$m$^2\cdot$ms$^{-1/2}$) \\ \hline
I. $T_{2a}>T_{2m}$, $D_a>D_m$          & 1.25                                 & 0.426                                        \\
II. $T_{2a}=T_{2m}$, $D_a=D_m$           & 1.30                                 & 0.502                                        \\
III. caliber variation only            & 1.45                                 & 0.450                                        \\
IV. undulation only            & 1.85                                 & 0.009                                        \\ \hline
\end{tabular}
\caption{Fit parameters of the time-dependent axial diffusivity $D(t)$ in our simulations in four axonal microgeometries (I-IV) specified in \figref{fig:ias-simulation-shape}a-b. 
}
\label{tab:ias-simulation-shape}
\end{table}

\begin{table}[th!]
\centering
\begin{tabular}{c|c c}
\hline
Dispersion angle $\theta$ ($^\circ$) & $D_\infty$ ($\mu$m$^2$/ms) & $c$ ($\mu$m$^2\cdot$ms$^{-1/2}$) \\ \hline
0                   & 1.25             & 0.426                                     \\
15                  & 1.16             & 0.407                                     \\
30                  & 0.94             & 0.343                                     \\
45                  & 0.62             & 0.253                                     \\ \hline
\end{tabular}
\caption{Fit parameters of the time-dependent axial diffusivity $D(t)$ in our simulations in 3{\it d}
(\figref{fig:ias-simulation-shape}e). Simulation results for dispersion angles $\theta$ = 15$^\circ$-30$^\circ$ are consistent with the human brain PGSE data in genu, cf. \tabref{tab:brain-pgse}.} 
\label{tab:ias-simulation-dispersion}
\end{table}

Numerical simulations for validating dMRI in brain microstructure (reviewed by ref. \cite{fieremans2018cookbook}) have been performed either in 2{\it d} or 3{\it d} simple geometries, 
or in combinations thereof. 
In particular, the axonal shape is typically modeled by artificial geometries. 
Recently, benefiting from the advances in microscopy, MC simulations were performed in $2d$ realistic microgeometry of neural tissue reconstructed from light microscopy \citep{chin2002simulation} or EM \citep{xu2018susceptibility},
and also in $3d$ realistic microstructure of astrocytes reconstructed from confocal microscopy \citep{nguyen2018gpu}. However, the crucial piece of the validation puzzle --- simulations in 3{\it d} realistic EM-based neuronal tissue microstructure (e.g., \figref{fig:ias-shape}a-b) --- have been  missing so far.

In {\it Methods, Monte Carlo simulation in realistic microstructure}, we describe our IAS segmentation and the MC simulations algorithm. 

To explore the possible cause of diffusion time-dependence along axons, we compare simulation results of four different microgeometries (\figref{fig:ias-simulation-shape}a): 

\renewcommand{\theenumi}{\Roman{enumi}}%
\begin{enumerate}
\item Original IAS segmentation from EM simulated with transverse relaxation time $T_{2a} = 80\,$ms and intrinsic diffusivity $D_a=2$ $\mu$m$^2$/ms in cytoplasm \citep{veraart2018teddi}  
and $T_{2m} = 20\,$ms and $D_m = 0.13$ $\mu$m$^2$/ms in mitochondria, assuming fully permeable mitochondria \citep{lopez1996mitochondria}.
This segmentation is the closest to reality and serves as the main result. 

\item The same IAS, but with no $T_2$ contrast and intrinsic diffusivity difference between mitochondria and axoplasm ($T_{2a} = T_{2m} = 80$ ms, $D_a=D_m=2$ $\mu$m$^2$/ms). 

\item Axially symmetric IAS with the same caliber variation (i.e., the same $z$-dependent cross-sectional area) as in the original IAS, but no undulation. 

\item IAS includes undulation and preserves volume, but has no caliber variation. The axonal skeleton describing the undulation is constructed by connecting the center of mass of each cross-section, and smoothed along the axon by a Gaussian filter of a standard deviation $\sigma$ = 1 $\mu$m. 
\end{enumerate}
All fibers are aligned to the $z$-axis, and the  orientation dispersion is not considered when comparing the above four cases. The effect of dispersion is considered separately in \figref{fig:ias-simulation-shape}e-h.

In the microstructure based on realistic IAS in \figref{fig:ias-simulation-shape}a (I-IV), the simulated overall $D(t)$ (from all axons) exhibits a notable time-dependence, which scales as $1/\sqrt{t}$ (\figref{fig:ias-simulation-shape}b). This is in agreement with our theoretical prediction of \eqref{eq:Dt-overall}, corresponding to the dynamical exponent $\vartheta= 1/2$ and the structural exponent of \eqref{p}, and confirms our expectations that the restrictions to diffusion along axons are due to short-range disorder. The corresponding bulk diffusivity $D_\infty$ and strength $c$ of restrictions (\eqref{eq:Dt-overall}) for all axons are listed in \tabref{tab:ias-simulation-shape}, based on \eqref{eq:D-c-overall} whereby individual axon's volume fraction $f_i$ and parameters ($D_{i,\infty}$, $c_i$) were obtained by fitting \eqref{eq:Dt-overall} to individual axon's $D_i(t)$.

The simulated $D(t)$ with or without considering low $T_2$ and low intrinsic diffusivity in mitochondria (I and II) shows similar diffusivity values and time-dependence (\figref{fig:ias-simulation-shape}b, \tabref{tab:ias-simulation-shape}). Similarly, compared with structure I, $D(t)$ of axially symmetric cylinders with only caliber variation (III) has slightly larger diffusivity values and very similar time-dependence. On the other hand, $D(t)$ of undulating fibers with no caliber variation (IV) shows much larger diffusivity values and negligible time-dependence ($\sim0.05$\% diffusivity change at $t=20-100$ ms), indicating that caliber variation is the main cause for the observed time-dependence. For the microgeometry I in \figref{fig:ias-simulation-shape}a
, the radius variation along individual axon, i.e. coefficient of variation of radii $\text{CV}(r)$, highly correlates with the relative diffusivity variation, i.e. $\zeta\equiv(D_0-D_{i,\infty})/D_{i,\infty}$ with the intrinsic diffusivity $D_0$, via a quadratic function (Pearson's $R$ = 0.8917 for $\zeta$ and $\text{CV}^2(r)$ in \figref{fig:ias-simulation-shape}d), a  relation derived in \eqref{eq:app-cv} in {\it Methods}. In the microgeometry I, $D_0$ is approximated by the volume-weighted sum of intrinsic diffusivities in IAS and mitochondria: $D_0 \simeq (1-f_m) D_a + f_m D_m \simeq 1.89\pm0.06\,\mu$m$^2$/ms, with the mitochondrial to IAS volume ratio $f_m\simeq6\%$ reported in Supplementary Fig. 1c.

\begin{figure*}[th!!]\centering
\includegraphics[width=0.96\textwidth]{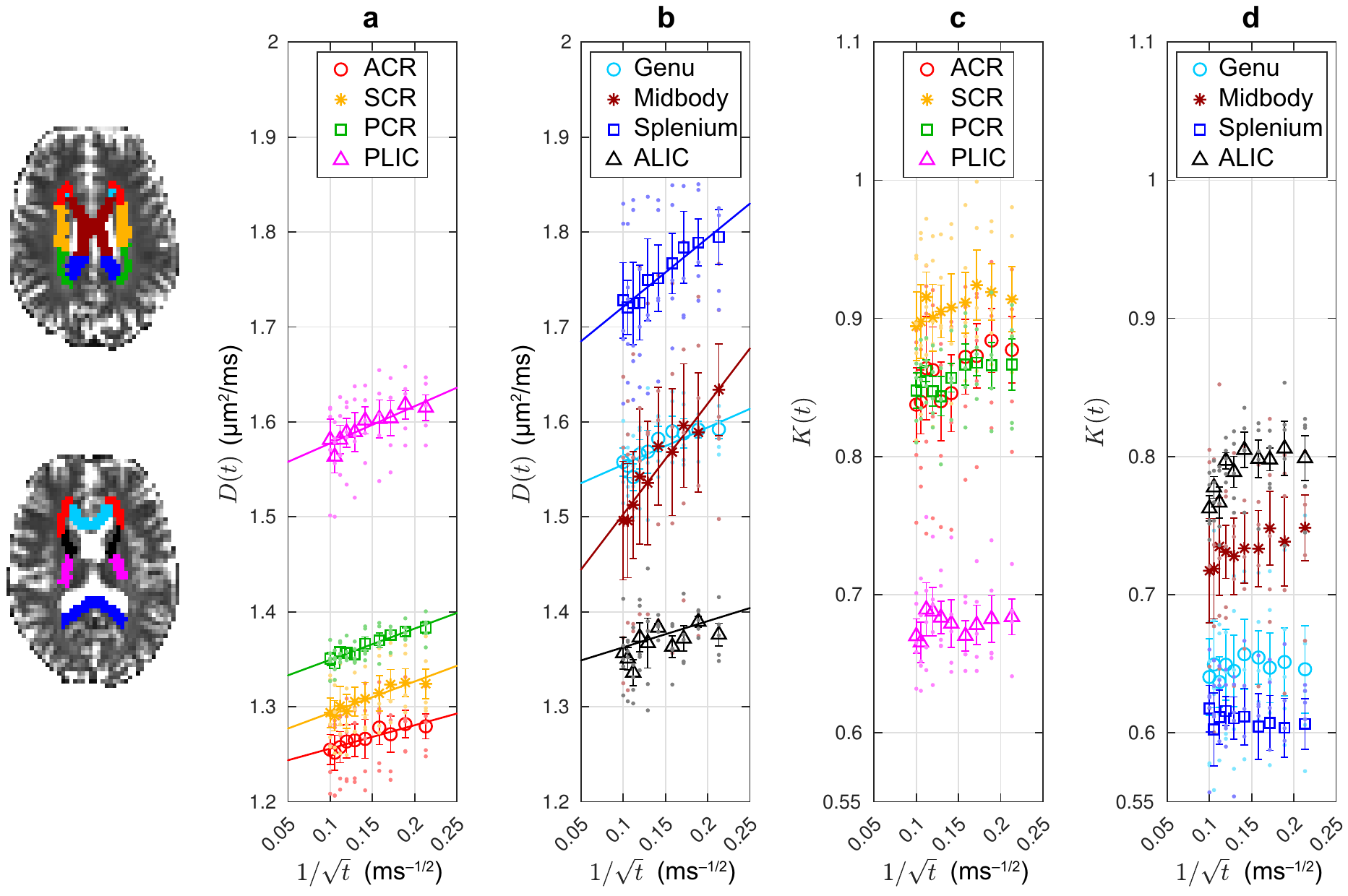}
\caption{Time-dependent axial diffusivity $D(t)$ ({\bf a}, {\bf b}) and axial kurtosis $K(t)$ ({\bf c}, {\bf d}) measured in vivo in brain WM of 5 healthy subjects using monopolar PGSE. In all WM ROIs, the axial diffusivity scales as $1/\sqrt{t}$ in \eqref{eq:Dt-overall} (P-value $<$ 0.05, \tabref{tab:brain-pgse}), confirming our prediction that the universality class along WM axons is $1d$ short-range disorder, cf. \figref{fig:ias-shape}. 
The fit parameters are summarized in \tabref{tab:brain-pgse}. 
\new{Points are plotted representing ROI-values of each subject, along with corresponding mean value (symbol and color in legend) and} error bars indicating the standard error over 5 subjects. (ACR/SCR/PCR = anterior/superior/posterior corona radiate, ALIC/PLIC = anterior/posterior limb of the internal capsule, genu/midbody/splenium of CC)}
\label{fig:3mm-AD}
\end{figure*}

It is essential to evaluate the {\it effect of fiber orientation dispersion} because the diffusion time-dependence transverse to individual axons could be projected to the main direction of the whole fiber bundle, confounding the $1/\sqrt{t}$-dependence in \eqref{eq:Dt-overall}. To evaluate this effect (\figref{fig:ias-simulation-shape}e-h), segmented axons in \figref{fig:ias-simulation-shape}a, scenario I, were oriented based on a Watson distribution with concentration parameters $\kappa$ = [$\infty$, 15.4, 4.7, 1.65] for cases of no dispersion up to high dispersion, corresponding to the overall polar dispersion angles $\theta$ = [0$^\circ$, 15$^\circ$, 30$^\circ$, 45$^\circ$], defined by $\theta\equiv\cos^{-1}\sqrt{\langle\cos^2\theta\rangle}$. \citep{novikov2018rotinv,lee2019em}
As a reference, the dispersion angle in the mouse brain corpus callosum \citep{lee2019em} is $\sim24^\circ$, corresponding to a $\kappa\sim6.9$.
This preserves the $D(t)$ scaling as $1/\sqrt{t}$, which overall decreases with increasing dispersion angle (\figref{fig:ias-simulation-shape}e), as manifested by the corresponding fit parameters in \eqref{eq:Dt-overall} (\tabref{tab:ias-simulation-dispersion}), bulk diffusivity in long-time limit and strength of restrictions: $D_\infty$ and $c\propto\langle\cos^2\theta\rangle$ (\eqref{eq:Dt-D-c-theta} and \figref{fig:ias-simulation-shape}g-h). In particular, the estimate of $c$ slightly deviates from this relation (\figref{fig:ias-simulation-shape}h) due to an extra $1/t$ term contributed by the diffusion transverse to individual axons, especially for the high dispersion case (large $\theta$, small $\langle\cos^2\theta\rangle$). Accounting for this small effect by using \eqref{eq:Dt-modify}, the corrected value of $c$ restores the relation.

The finite value of the time-dependence amplitude $c$ in \eqref{eq:Dt-overall} corresponds to about 4.4\% $D(t)$ change over $t = 20-100\,$ms time range. In particular, the axial diffusivity change $\propto\Delta(1/\sqrt{t})\sim\Delta t\cdot t^{-3/2}$ is even larger at short diffusion times. Including time-dependence for the intra-axonal compartment is therefore especially important for animal imaging \citep{sepehrband2016ghighg}, and for human dMRI \citep{huang2015highg} at relatively short diffusion times, achievable on high-gradient systems.

For the time-dependence of higher order cumulants of the intra-axonal signal, similar observation are made for the simulated overall kurtosis $K(t)$ in \figref{fig:ias-simulation-shape}c: In realistic IAS (\figref{fig:ias-simulation-shape}a, I and II) , $K(t)$ has almost the same values and overall $t$-dependence with or without considering low $T_2$ and low intrinsic diffusivity in mitochondria (\figref{fig:ias-simulation-shape}c). Similarly, compared with microgeometry I, the scenario with no axonal undulation (III) results in slightly smaller kurtosis values and similar $K(t)$ form. On the other hand, the scenario with no caliber variation (IV) shows much smaller kurtosis values and a totally different $K(t)$ form. These results indicate that the kurtosis time-dependence along realistic axons largely depends on caliber variation, rather than axonal undulation, with a small effect of low $T_2$ and low intrinsic diffusivity in mitochondria. For a fiber bundle with orientation dispersion, the simulated overall $K(t)$ increases with the dispersion angle (\figref{fig:ias-simulation-shape}f), especially for $\theta\gtrsim$ 30$^\circ$.

Focusing on the realistic microgeometry I without considering dispersion (blue data points in \figref{fig:ias-simulation-shape}c and \ref{fig:ias-simulation-shape}f), the simulated overall $K(t)$ ($\sim$0.4 at $t$ = 20-100 ms) consists of two parts: (1) the inter-compartmental contribution originating from the diffusivity differences between multiple axons (first RHS term in \eqref{eq:app-K}), and accounting for 24\% to 37\% of $K(t)$ at $t$ = 20-100 ms; and (2) the intra-compartmental contribution originating from individual axon's axial kurtosis (second RHS term in \eqref{eq:app-K}), and accounting for 76\% to 63\% of $K(t)$ at $t$ = 20-100 ms.

\begin{table}[t!!]
\small 
\centering
\begin{tabular}{l|ccc}
\hline
ROI      & P-value & $D_\infty$ ($\mu$m$^2$/ms) & $c$ ($\mu$m$^2\cdot$ms$^{-1/2}$) \\ \hline
ACR      & 6.3e-5  & 1.231 (0.005)             & 0.246 (0.034)                   \\
SCR      & 1.3e-5  & 1.261 (0.006)             & 0.330 (0.038)                   \\
PCR      & 3.3e-6  & 1.317 (0.005)             & 0.329 (0.033)                   \\
PLIC     & 1.2e-4  & 1.538 (0.010)             & 0.390 (0.062)                   \\
Genu     & 4.2e-4  & 1.516 (0.011)             & 0.391 (0.069)                   \\
Midbody  & 5.2e-6  & 1.386 (0.016)             & 1.17 (0.10)                    \\
Splenium & 1.9e-6  & 1.649 (0.009)             & 0.725 (0.058)                   \\
ALIC     & 1.8e-2  & 1.335 (0.020)             & 0.276 (0.129)                   \\ \hline
\end{tabular}
\caption{Fit parameters of the time-dependent axial diffusivity $D(t)$ in human brain data measured using monopolar PGSE (\figref{fig:3mm-AD}a-b). Standard errors are shown in the parenthesis. }
\label{tab:brain-pgse}
\end{table}

\subsection*{In vivo MRI demonstrates diffusion time-dependence along axons}

The time-dependent axial diffusivity $D(t)$, measured by monopolar PGSE in the human brain WM (\figref{fig:3mm-AD}a-b), were averaged over 5 healthy subjects ($n$ = 5) and plotted with respect to $1/\sqrt{t}$. In all studied WM ROIs, the axial diffusivity time-dependence demonstrates a $1/\sqrt{t}$ power-law relation in \eqref{eq:Dt-overall} (P-value $<$ 0.05, \tabref{tab:brain-pgse}), indicating that the universality class along human WM axons is short-range disorder (randomly distributed tissue inhomogeneity) in 1{\it d}, corresponding to a dynamical exponent $\vartheta$ = 1/2. The fit parameters in the different WM ROIs ($D_\infty$, $c$) are shown in \tabref{tab:brain-pgse}. \figref{fig:3mm-AD}c-d also shows that the axial kurtosis in WM ROIs from the same in vivo measurements is $\sim$0.8 and varies over diffusion time in some ROIs, demonstrating non-Gaussian diffusion along axons.  The data of 10 additional subjects scanned with higher resolution are in Supplementary Fig. 2.

\begin{figure}[bt!]\centering
\includegraphics[width=0.48\textwidth]{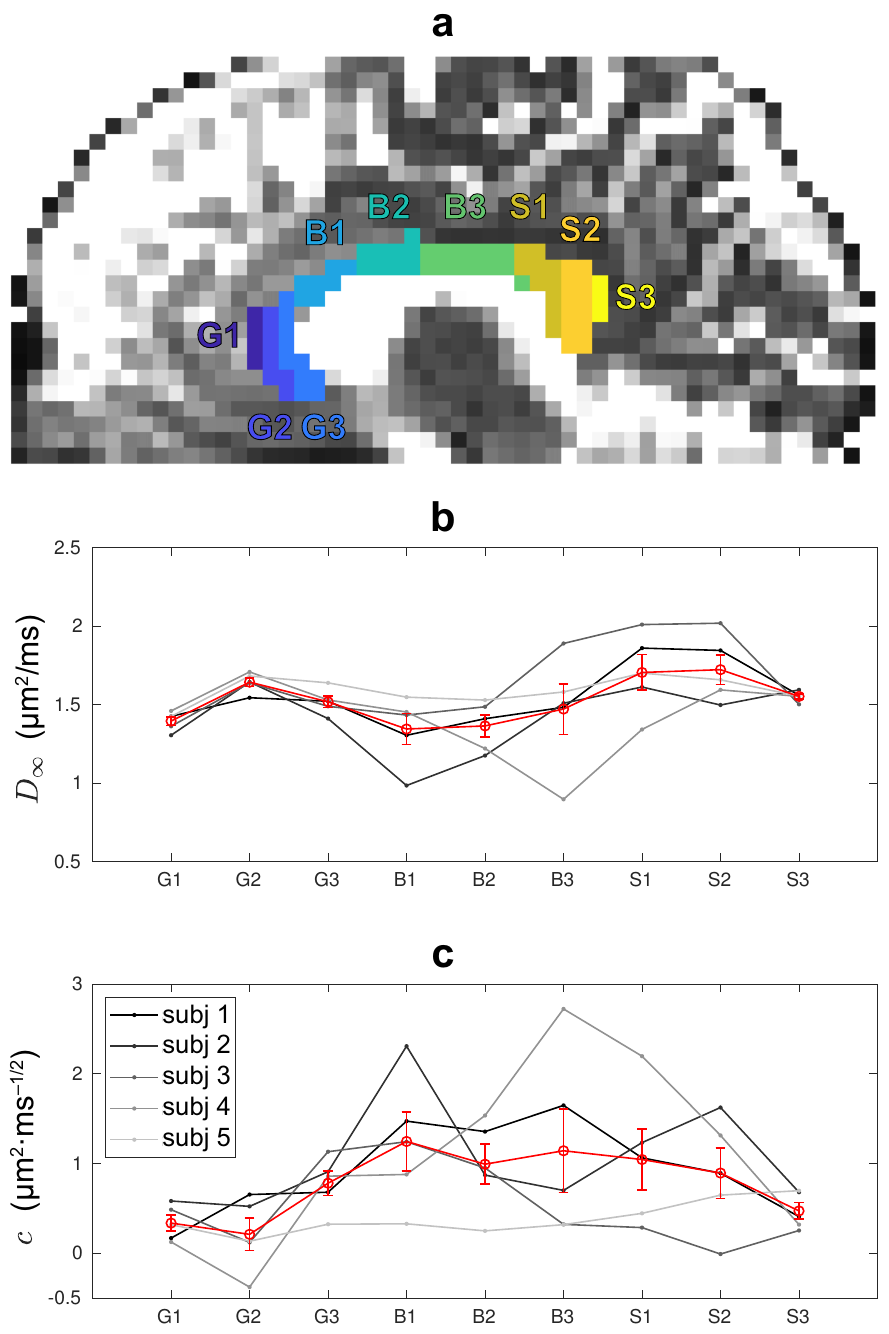}
\caption{Time-dependent parameters, $D_\infty$ and $c$ in \eqref{eq:Dt-overall}, estimated in vivo in brain CC of 5 healthy subjects using monopolar PGSE. {\bf a} Sub-regions of CC, including G1, G2, G3 for genu, B1, B2, B3 for midbody, and S1, S2, S3 for splenium. {\bf b} The bulk diffusivity $D_\infty$ 
shows a high-low-high pattern in CC. {\bf c} The strength $c$ of restrictions shows a low-high-low pattern in CC, indicating that the midbody has the largest distance between local caliber maxima \citep{fieremans2016invivo} and/or the widest bead shape, i.e. $c\propto\Gamma_{1d}|_{k_z\to 0,p=0}$ in \eqref{eq:power-spectrum-plateau}. The red data point is the mean of 5 subjects, and the error bar indicates the standard error of 5 subjects.}
\label{fig:cc-subregion}
\end{figure}

To further demonstrate the regional variation, \figref{fig:cc-subregion} shows the variation across the 9 sub-regions of corpus callosum (CC) (G1/G2/G3 for genu, B1/B2/B3 for midbody, and S1/S2/S3 for splenium) in the time-dependent parameters for each subject. The bulk diffusivity $D_\infty$ in $t\to\infty$ limit has a high-low-high pattern in genu-midbody-splenium in all subjects (\figref{fig:cc-subregion}b), whereas the strength $c$ of restrictions along axons has a low-high-low pattern in most of the subjects (\figref{fig:cc-subregion}c).

\subsection*{Time-dependent diffusion parameters alter in multiple sclerosis}

\begin{figure}[bt!]\centering
\includegraphics[width=0.48\textwidth]{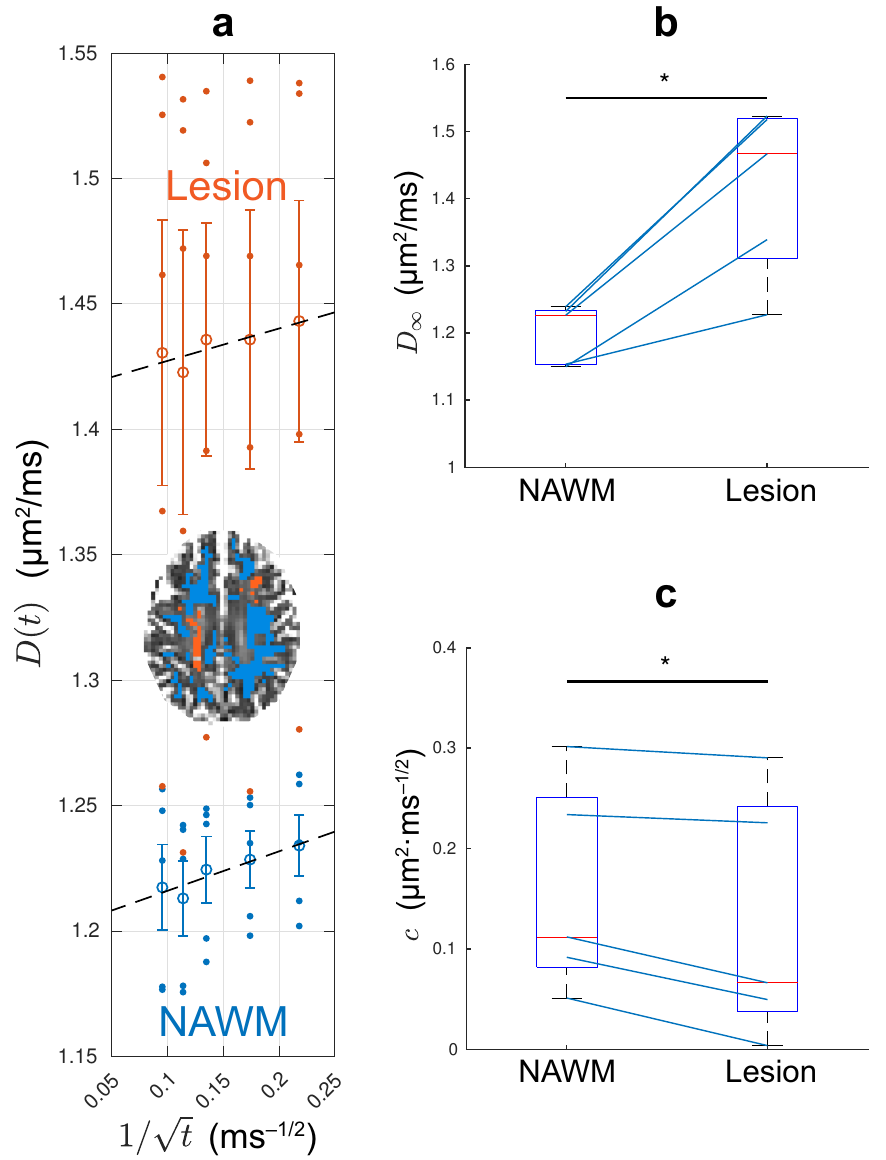}
\caption{Time-dependent axial diffusivity $D(t)$ and fit parameters ($D_\infty$, $c$) estimated in brain lesions (red) and normal-appearing white matter (NAWM, blue) of 5 MS patients using monopolar PGSE. {\bf a} \new{$D(t)$-values are plotted as points for each subject, along with corresponding mean value and error bars representing standard error over 5 subjects, and the corresponding fit to \eqref{eq:Dt-overall}, confirming the $1/\sqrt{t}$ functional form of $D(t)$.} {\bf b} The bulk diffusivity $D_\infty$ 
along axons is significantly larger in MS lesions than that in NAWM along axons. {\bf c} The corresponding strength $c$ of restrictions along axons is smaller in MS lesions than  in NAWM. In {\bf b} and {\bf c}, each patient is represented by a blue segment. The parameter differences between MS lesions and NAWM are compared by using one-sided Wilcoxon signed-rank test ($^*$P-value $<$ 0.05).}
\label{fig:ms}
\end{figure}

To evaluate the sensitivity of the time-dependent diffusion parameters to pathology, the time-dependent axial diffusivity $D(t)$ was measured by monopolar PGSE in multiple sclerosis (MS) lesions and normal appearing white matter (NAWM) in 5 MS patients ($n$ = 5). $D(t)$ averaged over subjects is plotted with respect to $1/\sqrt{t}$ in \figref{fig:ms}a, confirming that both in MS lesions and NAWM, $D(t)$ obeys the power-law relation in \eqref{eq:Dt-overall}, with P-values = 0.042 and 0.012 respectively.

The fit parameters ($D_\infty$, $c$, \figref{fig:ms}b-c) estimated individually in MS patients are compared between MS lesions and NAWM. The bulk diffusivity $D_\infty$ along axons in $t\to\infty$ limit is significantly larger in MS lesions than that in NAWM (P-value = 0.031, \figref{fig:ms}b).
Furthermore, the strength $c$ of restrictions along axons is significantly smaller in MS lesions than that in NAWM (P-value = 0.031, \figref{fig:ms}c).

\section*{Discussion}

The time-dependent diffusion MRI signal measured in vivo in brain white matter provides a signature for along-axon caliber variation. The specificity to this microstructural feature is determined here from a characteristic power-law decay of the diffusivity, and validated by performing realistic Monte Carlo simulations of diffusion inside axons from 3{\it d} EM images of mouse brain. In particular, our simulation results are consistent with in vivo measurements and the corresponding theoretical prediction that diffusion along axons is characterized by short-range disorder in 1{\it d}, with the dynamical exponent $\vartheta$ = 1/2 for \eqref{eq:Dt-overall}. This short-range disorder was confirmed by the power spectrum analysis of the actual shape of segmented myelinated axons in the 3{\it d} EM sample of mouse brain in this study, and was also observed in a preliminary study \citep{lee2020mchuman} performing MC simulations within realistic axons segmented from a large 3{\it d} EM sample of {\it human} subcortical WM.

Furthermore, simulations in different microgeometries based on this EM sample allow us to disentangle the contributions of different microstructural features to the overall 1{\it d} structural disorder, and reveal that the diffusivity and kurtosis time-dependence along axons is dominated by caliber variations, rather than axonal undulations.
For example, in {\it Supplementary Information}, simulations of diffusion in fiber bundles composed of fibers without caliber variations, such as undulation-only fibers (geometry IV in \figref{fig:ias-simulation-shape}a) or perfectly straight cylinders, demonstrate very small axial diffusivity time-dependence along the main direction, even for highly dispersed case (Supplementary Fig. 3).
Similarly, mitochondria have negligible impact on the time-dependence, due to their low volume fraction (Supplementary Fig. 1).
Yet, mitochondria are shown to correlate with axon caliber (\figref{fig:text-mito}), hence could indirectly impact the time-dependence, as discussed below for the MS pilot study.

Using PGSE dMRI in vivo in human brain, the  power-law scaling (\eqref{eq:Dt-overall}) was found in all WM ROIs (\figref{fig:3mm-AD}a-b). Particularly, in genu, the fit parameters of PGSE measurements (\tabref{tab:brain-pgse}) are of the same order as those in the MC simulations for dispersion angles $\theta$ = 15$^\circ$-30$^\circ$ (\tabref{tab:ias-simulation-dispersion}), which is consistent with the fiber dispersion $\approx$ 20$^\circ$ observed in histology \citep{ronen2014cc,lee2019em}.

The fitted power-law parameters show similar patterns over different WM regions between subjects (\figref{fig:3mm-AD}a-b), and especially in CC composed of highly aligned axons, demonstrating the potential of clinical applications in the future, as discussed later. 
Admittedly, the regional variations in the bulk diffusivity $D_\infty$ and strength $c$ of restrictions are noisy for individual subjects; however, we are still able to observe the general trends across the CC for the average over all subjects: The trends relate remarkably well to the pattern of axonal density in CC observed in histology \citep{aboitiz1992cc} and of axonal volume fraction in CC estimated via dMRI \citep{barazany2009axcaliber,fieremans2011dki}, as well as the higher spectrum of large axon diameters in the midbody according to ref. \citep{aboitiz1992cc}. On the one hand, the high-low-high trend in $D_\infty$ in CC could be related to the pattern of axonal density in CC observed in histology, with the assumption that the axial diffusivity in IAS is larger than that in extra-axonal space \citep{veraart2018teddi}. On the other hand, the low-high-low trend in $c$ in CC could be related with the bead width and/or distance between local caliber maxima along individual axons. \mpar{R2.1}\new{This observation cannot be supported or rejected by 2-dimensional histology and remains incompletely explained. For example, fiber bundles composed of (1) caliber-varying axons or (2) perfectly straight cylinders can have exactly the same 2-dimensional cross-sectional diameter distribution (Supplementary Fig. 3). Three-dimensional histology and analysis in different regions of CC are needed in the future to better understand our empirical observation of the trends across the CC.}

In addition to in vivo PGSE measurements in human brain WM reported here, the power-law dependence has also been reported using stimulated-echo (STE) measurements in vivo in human brain WM \citep{fieremans2016invivo} and ex vivo in spinal cord WM \citep{jespersen2018spinalcord}, where the diffusion time is varied by changing the mixing time. Both studies reported somewhat stronger time-dependence, as manifested by larger amplitude $c$ for the time-dependence (cf. Table 2 of ref. \citep{fieremans2016invivo} and Table 1 of ref. \citep{jespersen2018spinalcord} as compared to current study Tables \ref{tab:ias-simulation-dispersion} and \ref{tab:brain-pgse}), a potential overestimation caused by water exchange between intra-/extra-axonal water (fast diffusion, long $T_1$, $T_2$ values) and myelin water (slow diffusion, short $T_1$, $T_2$ values) during the mixing time of the STE sequence \citep{lee2017exchange}. Furthermore, in gray matter, the power-law dependence has been observed for the mean diffusivity using oscillating gradients in human brain \citep{baron2014ogse} and rat brain \citep{does2003ogse,novikov2014meso}, suggesting that the characteristics of short-range random  restrictions to diffusion along axons and dendrites are a universal feature of neuronal tissue.


Conventionally, diffusion in WM has been modeled using the featureless stick model (reviewed by ref. \cite{novikov2019review}), thereby assuming Gaussian diffusion, corresponding to a negligible axial intra-axonal kurtosis. Here, however, based on realistic simulations, combined with theory and experimental verification, we conclude that the intra-axonal axial kurtosis is non-negligible at clinical diffusion times. 
Indeed, for $t = 20-100\,$ms, the intra-axonal kurtosis along axons is $\sim$0.7 for $\theta$ = 30$^\circ$ based on simulations (\figref{fig:ias-simulation-shape}f), and $\sim$0.8 for  monopolar PGSE measurements in the human brain WM (\figref{fig:3mm-AD}c-d). The measured kurtosis in experiment is slightly larger than the intra-axonal kurtosis in simulations, likely due to the additional contribution of the extra-axonal space to the overall $K(t)$ in the measurement (\eqref{eq:app-K} in {\it Methods}).  

Simulations also demonstrated that the intra-axonal $K(t)$ increases with dispersion angle, especially for $\theta\gtrsim$ 30$^\circ$ (\figref{fig:ias-simulation-shape}f), which can be understood by the corresponding increasing range of intra-axonal diffusivity values when projected to the fiber bundle's main direction, resulting in a larger contribution to the overall $K(t)$, i.e. the first right-hand-side term in \eqref{eq:app-K}. Hence, the higher order cumulants of the intra-axonal signal, including $K$, are very sensitive to the fiber dispersion (i.e., the functional form and the degree of orientation distribution), and should  be incorporated in future biophysical models of dMRI in WM.

Besides the nominal (nonzero) value of the axial kurtosis, the observed time-dependence of both $D(t)$ and $K(t)$ are non-trivial and should be considered in WM biophysical modeling. For $D(t)$ along axons, proportional to $\Delta(1/\sqrt{t})\sim\Delta t\cdot t^{-3/2}$, it is negligible only when the time range $\Delta t$ is small (e.g., $\Delta t<$ 5 ms), or the diffusion time is long (e.g., $t>$ 200 ms). For $K(t)$, our simulations in IAS show 7\% changes over the clinical time range $t$ = 20-100 ms.

The observation of axon caliber variation and beading with non-invasive time-dependent diffusion MRI calls for evaluating the role of this microstructural feature in pathology. In this work, we demonstrated altered diffusion time-dependence along axons in WM lesions versus NAWM of 5 MS patients (\figref{fig:ms}), with corresponding changes in the fit parameters that are potentially related to specific pathological changes. In particular, the increase in the bulk diffusivity $D_\infty$ along axons in MS lesions versus NAWM (\figref{fig:ms}b) may suggest ongoing demyelination and axonal loss \citep{moll2011ms}. While this observation alone has been reported before with dMRI \citep{werring1999ms,guo2001ms,mustafi2019ms}, our results reveal, for the first time to our knowledge, that the diffusivity time-dependence along WM axons, i.e. $c$ in \eqref{eq:Dt-overall}, is smaller in MS lesions than that in NAWM, with $c\propto$ the correlation length $l_c$. \citep{fieremans2016invivo} This observation is potentially indicative of an increase in mitochondria density, a feature of chronic demyelination documented from histology in axons and astrocytes of WM lesions \citep{witte2014msmito}. Since mitochondria and axon caliber are shown to correlate (\figref{fig:text-mito}c) \citep{wang2003varicosity}, an increase in mitochondria would shorten the correlation length $l_c$ that characterizes the distance between local maxima of the axon. Hence, the parameter $c$ potentially targets the specific pathology of mitochondria increase in MS. 

Conventional MRI methods (e.g., $T_2$-FLAIR) are well-known to distinguish MS lesions from NAWM, and typically attributed to {\it demyelination} \citep{witte2014msmito}. Here, however, we aim to use MS lesion data to {\it in vivo} validate the strength $c$ of restrictions in \eqref{eq:Dt-overall} as a specific measure for changes in {\it mitochondria}: We demonstrate significant difference in $c$ between MS lesions and NAWM (\figref{fig:ms}c), and attribute to an increase in mitochondria as a response to demyelination in MS lesions \citep{witte2014msmito}. This observation may contribute to understanding the underlying pathological mechanisms taking place in MS lesion formation. In addition, our finding also suggests diffusion time-dependence measurements as potential biomarker suitable for monitoring other pathologies presenting increased neurite beadings due to other mechanisms (rather than mitochondrial increase).

In addition to MS \citep{trapp1998ms}, axonal beading in WM has been observed in several other pathologies, such as traumatic brain injury (TBI) \citep{tang2012microtubule,johnson2013tbireview}, and ischemic stroke \citep{garthwaite1999beadischemia}.

Axonal varicosities, or axonal beading along axons, can be a pathological change caused by accumulation of transported materials in axonal swellings after TBI \citep{tang2012microtubule,johnson2013tbireview}; it has been observed that varicosities arise during dynamic stretch injury, caused by microtubule breakdown and partial transport interruption along axons. Furthermore, varicosities due to ischemic injury to WM axons can be caused by Na\textsuperscript{+} loading of the axoplasm, which leads to a lethal Ca\textsuperscript{+} overload through reversed Na\textsuperscript{+}-Ca\textsuperscript{+} exchange \citep{garthwaite1999beadischemia}. Hence, the average distance between varicosities is potentially a biomarker for axonal injury in TBI and ischemia, facilitating evaluation of the effectiveness of treatment and rehabilitation services. Since the average distance between varicosities along axons is of the order of $10\,\mu$m \citep{garthwaite1999beadischemia,tang2012microtubule,johnson2013tbireview}, much smaller than the resolution of most of the clinical imaging techniques, dMRI is the method of choice to estimate in vivo the pathological change of TBI  \citep{inglese2005tbidti,kinnunen2010tbidti} 
and of ischemic stroke \citep{moseley1990strokecat}. In particular, time-dependent diffusion tensor imaging may enable the estimation of the correlation length of varicosities along axons, related to the average distance between varicosities, a potential biomarker for monitoring TBI and ischemic stroke patients.

Besides beading in WM, the ubiquitous $1/\sqrt{t}$ time-dependence along neurites in gray matter \citep{does2003ogse,novikov2014meso,papaioannou2017gm} suggests possible applications in other neurodegenerative diseases. For instance, reduced density of axonal varicosities was observed in the human superior frontal cortex of mild Alzheimer disease \citep{ikonomovic2007ad}; decreased dendritic spine density was observed in the human prefrontal cortex of Schizophrenia \citep{glantz2000schizophrenia}; and an increased density of axonal varicosities was observed in injured dopaminergic neurons in the rat substantia nigra, an animal model of Parkinson's disease \citep{finkelstein2000pd}. The ability to evaluate restriction changes along neurites opens a door to monitoring the progression and therapy response of these diseases.


Thanks to recent advances in $3d$ EM \citep{baena2019em}, our works for the first time to our knowledge demonstrates the feasibility to employ EM-derived microstructure as numerical phantoms for realistic $3d$ simulations.  
By fully controlling the microgeometry of numerical phantoms, MC simulations provide complete flexibility to evaluate the influence of different microstructural features. Here they were employed to elucidate that the time-dependent diffusion signal along axons mainly originates from the caliber variations, with the contributions from mitochondria and axonal undulations having relatively small effects. 

The value of realistic simulations as a validation tool already being clear, and one can think of extending this approach further to study the sensitivity of MRI to microstructure. Larger EM samples \citep{wetzel2016largeem} would be needed to enable diffusion simulations at longer diffusion times. For the EM sample used in the current study, the maximal axon length $L\sim 18\,\mu$m corresponds to a length-related correlation time $\tau_L= L^2/(2D_0) \simeq 80\,$ms for $D_0 \sim 2\,\mu$m$^2$/ms, which sets the maximum feasible diffusion time for the simulation. Furthermore, we only focused on the intra-axonal geometry of myelinated axons in WM. 
Although, the contribution of extra-axonal space is non-negligible, extra-axonal signals are relatively smaller than the intra-axonal ones because of the shorter $T_2$ in extra-axonal space and long echo time applied in experiments \citep{veraart2018teddi}. For the diffusivity time-dependence along the fiber bundle, we expect that the diffusivity time-dependence in extra-axonal space is similar to that in intra-axonal space, since water molecules experience similar beading arrangement in either intra- or extra-axonal spaces.
Faithfully segmenting and simulating the diffusion in the extra-axonal space is needed to understand how robust the observed power-law is with increasing dispersion.  In addition, other structures, such as unmyelinated axons, glia cell and blood vessels, may have nontrivial contributions to the (time-dependent) diffusion signal and can be added to the numerical microgeometry.
Ultimately, large human EM sample \citep{lee2020mchuman} (in a scale of MRI voxel size), prepared with extra-cellular space preserving technique if possible, would provide so far the most representative numerical phantom to the human tissue microstructure after fully segmenting all the cells inside the sample.

Finally, while the proposed framework here focuses on performing MC to model diffusion in realistic WM microstructure, it can also be applied to gray matter, or tissue samples with pathology. In addition, the framework can be extended to include other MR contrast mechanisms, e.g., magnetization transfer, mesoscopic susceptibility \citep{Kiselev2018}, 
$T_1$ and $T_2$ relaxation \citep{Does2018}, and water exchange \cite{mukherjee2016exchange}, thereby facilitating the exciting ability to validate non-invasive MRI.

\section*{Methods}
All procedures performed in studies involving animals were in accordance with the ethical standards of New York University School of Medicine. All mice were treated in strict accordance with guidelines outlined in the National Institutes of Health Guide for the Care and Use of Laboratory Animals, and the experimental procedures were performed in accordance with the Institutional Animal Care and Use Committee at the New York University School of Medicine. 
All procedures performed in studies involving human participants were in accordance with the ethical standards of New York University School of Medicine. All protocols were approved by the local institutional review board (New York University School of Medicine). Informed consent was obtained from all individual participants included in the study.

\subsection*{Electron microscopy and intra-axonal space segmentation} \label{sec:methods-em}
The brain tissue from a female 8-week-old C57BL/6 mouse's genu of corpus callosum (CC) was processed and analyzed with a scanning electron microscope (SEM) (Zeiss Gemini 300 SEM with 3View). Part of the data was discarded due to unstable quality, leading to a volume (\figref{fig:ias-shape}a) of 36$\times$48$\times$20 $\mu$m$^3$. To segment long axons passing through all slices, we employed a simplified seeded region growing algorithm \citep{adams1994srg,lee2019em,abdollahzadeh20193dem}. The segmented axons (\figref{fig:ias-shape}b) shorter than 20 $\mu$m were discarded, leading to 227 long axons ($\geq$ 20 $\mu$m in length). More details were reported in our previous work \citep{lee2019em}.

The intra-axonal space (IAS) segmentation was down-sampled into a voxel size of (0.1 $\mu$m)$^3$. The effect of orientation dispersion was controlled by subsequently realigning axons along the $z$-axis (\figref{fig:ias-simulation-shape}a). The aligned axons were truncated at both ends by 1 $\mu$m to avoid oblique end faces, resulting in axons of about 18 $\mu$m in length.

\begin{figure}[t!!]\centering
\includegraphics[width=0.48\textwidth]{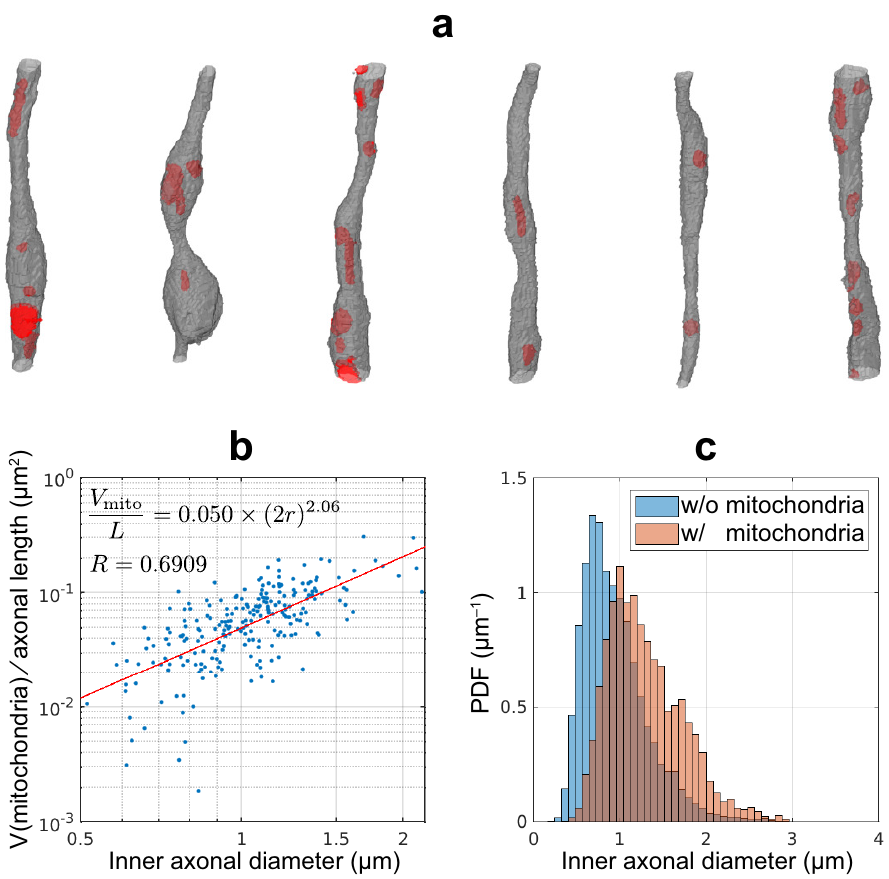}
\caption{ Mitochondria segmentation and characterization of the relation between mitochondria and axon caliber or diameter: {\bf a} $\sim$1,300 mitochondria (red) in the 227 axons (gray) of \figref{fig:ias-shape}a-b were manually segmented. {\bf b} The inner axonal diameter $r$ approximately correlates with each axon's mitochondrial volume per unit length ($V_\text{mito}/L$) via a quadratic function. {\bf c} Axonal diameters in cross-sections where mitochondria are present (red) are significantly larger, compared to cross-sections where mitochondria are absent (blue) (P-value $<$ 0.001). }
\label{fig:text-mito}
\end{figure}

\subsection*{Mitochondria density affects inner axonal diameter}

To evaluate the influence of mitochondria on the axon caliber variation, and on the diffusion time-dependence, we manually segmented $\sim$1,300 mitochondria in 227 axons (\figref{fig:text-mito}a). For individual axons, their inner diameter is found to correlate with the mitochondrial volume per unit length via a quadratic function (\figref{fig:text-mito}b), similar to the observation in ref. \citep{perge2009mitochondria}. In addition, the axonal diameters calculated based on cross-sections with and without the presence of mitochondria are 1.29 $\pm$ 0.43 $\mu$m ($n$ = 13,653) and 0.94 $\pm$ 0.38 $\mu$m ($n$ = 31,747), respectively (\figref{fig:text-mito}c), indicating that the presence of mitochondria in IAS corresponds to larger axonal diameters (P-value $<$ 0.001), and in agreement with a previous histological study in human and non-human primate retinas \citep{wang2003varicosity}.

\subsection*{Monte Carlo simulation in realistic microstructure}

MC simulations of random walkers were implemented in CUDA C++ for diffusion in a continuous space. 2.27$\times$10$^9$ walkers in total were employed inside 3{\it d} segmentations of 227 IASs, with 1$\times$10$^7$ walkers per IAS. The walker encountering the cell membrane is elastically reflected or permeates through the membrane based on a permeation probability for highly permeable membranes \cite{baxter2013simulation}, $P_{1\to2}=\text{min}(1,\sqrt{D_2/D_1})$ with intrinsic diffusivities $D_1$ and $D_2$ in compartments 1 and 2.
The top and bottom faces of each IAS binary mask, artificially made due to the length truncation, were extended with its reflective copies (mirroring boundary condition) to avoid geometrical discontinuity in diffusion simulations \citep{fieremans2018cookbook}.

Each particle diffused over 5$\times$10$^5$ steps with a duration $\delta t$ = 2$\times$10$^{-4}$ ms and a length $\sqrt{6D_a \delta t}$ = 0.049 $\mu$m for each step in IAS and $\sqrt{6D_m \delta t}$ = 0.013 $\mu$m in mitochondria, where the intrinsic diffusivity, $D_a$ = 2 $\mu$m$^2$/ms in IAS and $D_m$ = 0.13 $\mu$m$^2$/ms for each step in mitochondria, is taken to agree with recent in vivo experiments \citep{novikov2018rotinv,veraart2019highb} and previous in vitro study \citep{lopez1996mitochondria}. Maximal diffusion time in simulations is $t$ = 100 ms. Total calculation time was $\sim$4 days on $\sim$20 NVIDIA Tesla V100 GPU on the NYU Langone Health BigPurple high-performance computing cluster.

The $i$-th axon's moment tensors $\langle x_{j_1} x_{j_2}\rangle_i$ and $\langle x_{j_1} x_{j_2} x_{j_3} x_{j_4}\rangle_i$ are calculated based on the simulated diffusion displacement vector $\bf{x}$ (with the component $x_{j_1}$, $j_1$ = 1, 2, or 3) \citep{jensen2005dki,jensen2010dki}, and their projections yield the axon's apparent diffusivity $D_i(t,\hat{\bf n})$ and apparent kurtosis $K_i(t,\hat{\bf n})$ in the direction $\hat{\bf n}$ (with the component $n_{j_1}$) \citep{jensen2005dki,jensen2010dki}:
\begin{subequations} \begin{align}
D_i(t,\hat{\bf n})&=\frac{\langle s^2\rangle_i}{2t}\,,\\
K_i(t,\hat{\bf n})&=\frac{\langle s^4\rangle_i}{\langle s^2\rangle_i^2}-3\,,
\end{align}\end{subequations} \addtocounter{equation}{-1}
where
\begin{subequations} \begin{align*}
\langle s^2\rangle_i &= n_{j_1}n_{j_2}\langle x_{j_1}x_{j_2}\rangle_i\,,\\
\langle s^4\rangle_i &= n_{j_1}n_{j_2}n_{j_3}n_{j_4}\langle x_{j_1}x_{j_2}x_{j_3}x_{j_4}\rangle_i\,,
\end{align*} \end{subequations}
and the summation over the pairs of repeating indices is implied.

For an axon oriented into the direction $\hat{\bf n}'$, the axon's axial diffusivity and axial kurtosis defined along the $z$-axis direction $\hat{\bf z}$ were calculated based on the estimated moment tensors, yielding $D_i(t,\hat{\bf n})$ and $K_i(t,\hat{\bf n})$, where $\hat{\bf n}=2(\hat{\bf n}'-\hat{\bf z})\hat{\bf z}-\hat{\bf n}'$. This is similar to the reflection of light, with the incident light along $-\hat{\bf n}'$ falls on the surface normal to $\hat{\bf z}$, and is reflected along $\hat{\bf n}$. It is then straightforward to calculate the overall $D(t)$ and $K(t)$ using \eqref{eq:app-D-K} below.

\subsection*{Ensemble averaging over axons}
\label{sec:methods-theory}

The dMRI signal from many axons can be approximated by the cumulant expansion \citep{jensen2005dki,kiselev2017mrphysics}
\begin{eqnarray} \notag 
S(b,t) &\simeq& e^{ -bD(t)+\frac{1}{6}b^2 D^2(t) K(t) + {\cal O}(b^3) }\\ \label{eq:app-signal}
&=& \sum_i^{} f_i \cdot e^{ -bD_i(t)+\frac{1}{6}b^2 D_i^2(t) K_i(t) + {\cal O}(b^3) }\,,
\end{eqnarray}
where $D(t)$ and $K(t)$ are overall diffusivity and kurtosis, and $D_i(t)$ and $K_i(t)$ are diffusivity and kurtosis of individual axons with volume fractions $f_i$, such that $\sum_i f_i \equiv 1$. 
Expanding  \eqref{eq:app-signal} up to $b^2$, we obtain \citep{jensen2005dki,dhital2018restricted}
\begin{subequations} \label{eq:app-D-K} \begin{align} \label{eq:app-D}
D(t)&= \langle D_i(t)\rangle \equiv \sum_i^{} f_i\cdot D_i(t)\,,\\ \label{eq:app-K}
K(t)&=\frac{1}{D^2(t)} \sum_i^{} \left[ 3f_i\cdot(D_i(t)-D(t))^2 + f_i\cdot D_i^2(t) K_i(t)\right].
\end{align} \end{subequations}
Equation~(\ref{eq:app-D}) yields that the overall $D_\infty$ and $c$ entering \eqref{eq:Dt-overall} are given by the volume-weighted averages of the corresponding parameters of the individual axons, 
\begin{equation} \label{eq:D-c-overall}
D_\infty\equiv \langle D_{i,\infty}\rangle\,,\quad c\equiv\langle c_i\rangle\,.
\end{equation}

Throughout, we use the time interval $t = 20-80\,$ms to fit $D_{i,\infty}$ and $c_i$ from MC simulations of individual axons, i.e. fitting \eqref{eq:Dt-overall} to $D_i(t)$, and  employ these parameters to predict the axial diffusivity $D(t)$ of all axons in \eqref{eq:Dt-overall} and \eqref{eq:D-c-overall}. The maximal diffusion time used for fitting is bounded by the axonal length of the EM substrate $L\sim 18 \,\mu$m:  $L^2/(2D_0)\simeq 80\,$ms for $D_0=2\,\mu$m$^2$/ms.

Considering a fiber bundle with the orientation dispersion, the diffusion displacement within an axon (dispersed into $\theta_i$) is generally along the axon due to its thin size. Its projection to the fiber bundle's main direction leads to a contribution to the second order cumulant $\langle s^2\rangle_i\propto\cos^2\theta_i$ along the fiber bundle. As a result, the overall diffusivity and corresponding parameters are given by
\begin{equation} \label{eq:Dt-D-c-theta} 
\frac{D(t)}{D(t)|_{\theta=0}}=\frac{D_\infty}{D_{\infty}|_{\theta=0}} = \frac{c}{c|_{\theta=0}}=\langle\cos^2\theta\rangle\,.
\end{equation}
However, for a highly dispersed fiber bundle (e.g., $\theta$ = 45$^\circ$), some axons are oriented roughly perpendicular to the fiber bundle's main direction; these axons' radial diffusivity $\propto 1/t$ can be projected to the main direction, resulting in a small contribution to the overall axial diffusivity $D(t)$ \citep{fieremans2016invivo}, biasing the estimate of $c$. To account for this contribution, a correction term is added to the overall $D(t)$ in \eqref{eq:Dt-overall}:
\begin{equation} \label{eq:Dt-modify}
D(t)\simeq D_\infty + c\cdot\frac{1}{\sqrt{t}} + c'\cdot\frac{1}{t}\,,
\end{equation}
where $c'$ is related with caliber variation \citep{burcaw2015meso} and undulation \citep{lee2019radial}.

\subsection*{Relation of relative caliber variation and relative diffusivity variation}
\label{sec:app-cv}

In \figref{fig:ias-simulation-shape}d, the metric specifying the axonal shape, the coefficient of variation of radius $\text{CV}(r)=\delta r / \langle r\rangle$, where $\delta r$ is the standard deviation and $\langle r \rangle$ is the mean radius, and the relative diffusivity variation $\zeta\equiv(D_0-D_{i,\infty})/D_{i,\infty}$ \citep{novikov2011rpbm} highly correlate with each other. Interestingly, $\text{CV}(r)$ is calculated solely based on axons' 3{\it d} microgeometry; in contrast, the relative diffusivity variation  is estimated based on simulation results. To explain this observation, we derive  a simple relation to link the two metrics.

Our argument is based on the coarse-graining of $1d$ axonal microstructure by diffusion \citep{novikov2014meso,novikov2019review}.
When the diffusion length $L_d(t)$ grows beyond the correlation length of caliber variations, all the effectively $1d$ diffusion physics is represented by a one-dimensional coarse-grained diffusion coefficient $D(z)$ varying in space on the scale $L_d(t)$. 
For sufficiently large $L_d(t)$ (long $t$), the 
local fluctuations $\delta D(z) = D(z)-\overline{D}$ become small, $|\delta D(z)| \ll \overline{D}$, where $\overline{D}$ is the average of $D(z)$ along the axon.
In particular, the local fluctuation of the coarse-grained local $1d$ diffusivity $\delta D(z)\simeq (\partial\overline{D}/\partial\bar{n})\delta n(z)$ is proportional to the local fluctuation of restriction density $\delta n$, with $\bar{n}$ the mean density \citep{novikov2014meso}. It is then straightforward to calculate each individual axon's bulk diffusivity $D_{i,\infty}$, given by \citep{novikov2011rpbm}
\begin{eqnarray*}
\frac{1}{D_{i,\infty}} &=& \left\langle \frac{1}{D_{i}(z)} \right\rangle_z
\simeq \frac{1}{\, \overline{D} \,}\left[1+\frac{\langle (\delta D)^2\rangle_z}{\overline{D}^2}\right]\\
&\simeq& \frac{1}{\,\overline{D}\,}\left[ 1+\left( \frac{\partial \ln \overline{D}}{\partial\bar{n}}\right)^2 \langle (\delta n)^2 \rangle_z \right],
\end{eqnarray*}
simplified as
\begin{equation} \label{eq:app-n2}
\frac{\overline{D}-D_{i,\infty}}{D_{i,\infty}}\propto \langle (\delta n)^2\rangle_z
\end{equation} 
to the lowest order in $\delta n$. Above, we neglected the third and higher orders of $\delta n$, and so this derivation is by construction perturbative, valid for small 
$\zeta$ and $\text{CV}(r)$. 

The cross-sectional area (CSA) variation $A(z)$ along an axon can be expressed as the convolution of restriction density $n(z)$ and shape function of a restriction $v(z)$, i.e. $A(z)= n(z)\ast v(z)$, or in the Fourier domain, $A(k_z) = n(k_z) v(k_z)$. 
The coarse-grained density fluctuation $\delta n(k_z)$ at scales much longer than the mean restriction width $\bar{l}$, 
corresponding to $k_z \cdot \bar{l} \ll 1$, causes the corresponding fluctuation 
$$
\delta A(k_z) = \delta n(k_z) v(k_z) \simeq \delta n(k_z) v_0 \,, \quad v_0=v|_{k_z=0}\sim\overline{A}\cdot\bar{l} \,.
$$ 
Here $v_0$ is the restriction power (e.g., single bead volume). 
Hence, $\delta A(k_z)/\overline{A} \sim \delta n(k_z) \bar{l}$, or 
$$
\delta n(z) \propto \delta A(z)/\overline{A} \propto \delta r(z)/\langle r\rangle,
$$ 
since 
$\delta A \sim \langle r \rangle \cdot \delta r$. 
Substituting into \eqref{eq:app-n2}, and approximating the local average diffusivity by the free diffusivity $\overline{D}\simeq D_0$ (no restrictions for $\delta n =0$), we obtain
\begin{equation}\label{eq:app-cv}
\zeta\equiv \frac{D_0-D_{i,\infty}}{D_{i,\infty}}
\propto \frac{\langle (\delta A)^2\rangle}{\overline{A}^2}
\propto \frac{\langle (\delta r)^2\rangle}{\langle{r}\rangle^2} =\text{CV}^2(r)\,,
\end{equation} 
which is demonstrated by plotting $\zeta$ versus $\text{CV}^2(r)$ in \figref{fig:ias-simulation-shape}d, where the correlation coefficient = 0.8917 is high, and a small intercept = 0.09 verifies this simple relation. We note that \eqref{eq:app-cv} has been derived for small $\zeta$ and $\text{CV}(r)$, and the scatter close to the origin in \figref{fig:ias-simulation-shape}d is indeed much closer to the straight line.

\subsection*{In vivo MRI of healthy subjects}
dMRI measurements were performed on 5 healthy subjects (4 males/1 female, 21-32 years old) using a monopolar PGSE sequence provided by the vendor (Siemens WIP 919B) on a 3T Siemens Prisma scanner (Erlangen Germany) with a 64-channel head coil. For each subject, we varied the diffusion time $t$ = [22, 28, 34, 40, 50, 60, 70, 80, 90, 100] ms and fixed the diffusion gradient pulse width $\delta$ at 15 ms. For each scan, we obtained three $b$ = 0 non-diffusion weighted images and 62 diffusion weighted images (DWIs) of b-values $b$ = [0.4, 1, 1.5] ms/$\mu$m$^2$ along [12, 20, 30] gradient directions for each b-shell, with an isotropic resolution of (3 mm)$^3$ and a field-of-view (FOV) of 210$\times$204 mm$^2$. The whole brain volume was scanned within 30 slices, aligned parallel to the anterior commissure(AC)-posterior commissure (PC) line. GRAPPA with acceleration factor = 2 and multiband with acceleration factor = 2 were used. All scans were performed with the same TR/TE = 4000/139 ms. Total acquisition time is $\sim$60 min for each subject. In the main text, we focus on this dataset. The data of 10 additional subjects scanned with a smaller voxel size, exhibiting similar outcomes, are shown in {\it Supplementary Information}.

Our image processing DESIGNER pipeline is based on ref. \citep{ades2018designer} and includes 5 steps: denoising, 
Gibbs ringing elimination, 
eddy-current and motion correction, 
and Rician noise correction. 

For each voxel, we fitted dMRI data to the diffusion and kurtosis tensor using weighted linear least square (WLLS) \citep{veraart2013wlls}, and calculated eigenvalues of the diffusion tensor (in the order of $\lambda_1 \geq \lambda_2 \geq \lambda_3$) and the fractional anisotropy (FA) accordingly \citep{basser1994dti}. Experimental axial diffusivity is defined by $D\equiv\lambda_1$, and experimental axial kurtosis is defined by the apparent kurtosis along the principal axis of the diffusion tensor.

Each subject's mean FA map, averaged over all diffusion time points, was registered to FSL's standard FA map with FMRIB's linear and non-linear registration tools (FLIRT, FNIRT) \citep{jenkinson2001flirt,andersson2007fnirt}. We retrieved the transformation matrix (FLIRT) and the warp (FNIRT) to inversely transform John's Hopkins University (JHU) DTI-based WM atlas ROIs \citep{mori2005atlas} to the individual space. Cerebrospinal fluid (CSF) mask was segmented by FSL, FAST \citep{zhang2001fast} and expanded by 1 voxel to exclude WM voxels close to CSF. We focused on main WM tracts, such as anterior corona radiata (ACR), posterior corona radiata (PCR), superior corona radiata (SCR), anterior and posterior limb of the internal capsule (ALIC, PLIC), genu, midbody, and splenium of the corpus callosum.

To further discuss the variation of tissue properties in CC, we divided CC ROIs defined in JHU DTI atlas into 9 sub-regions in total (\figref{fig:cc-subregion}a), such as G1, G2, G3 for genu, B1, B2, B3 for midbody, and S1, S2, S3 for splenium. The 9 sub-regions are then co-registered and transformed to individual subject's space by using FSL.

\subsection*{In vivo MRI of multiple sclerosis patients}
The dMRI measurements were performed on 5 MS patients (5 females, 32-48 years old) using a monopolar PGSE sequence provided by the vendor (Siemens WIP 511E) on a 3T Siemens Prisma scanner (Erlangen Germany) with a 64-channel head coil. For each subject, we varied the diffusion time $t$ = 21-110 ms and fixed the diffusion gradient pulse width $\delta$ at 15 ms. For each time point, we obtained three $b$ = 0 non-diffusion weighted images and DWIs of $b$ = 0.5 ms/$\mu$m\textsuperscript{2} along 30 gradient directions, with an isotropic resolution of (3 mm)$^3$ and an FOV of 222$\times$222 mm$^2$. A slab of the brain volume was scanned within 15 slices, aligned parallel to the AC-PC line. All scans were performed with the same TR/TE = 4200/150 ms. Total acquisition time of DWIs is $\sim$15 min for each subject. 

Sagittal 3{\it d} MPRAGE brain images were acquired with an isotropic resolution of (1 mm)\textsuperscript{3}, an FOV of 256$\times$256 mm\textsuperscript{2}, TR/TE = 2100/2.72 ms, and inversion time = 900 ms. Axial FLAIR brain images were acquired with an anisotropic resolution of 0.6875$\times$0.6875$\times$5 mm\textsuperscript{3}, an FOV of 220$\times$220 mm\textsuperscript{2}, TR/TE = 9000/90 ms, and inversion time = 2500 ms.

The image processing pipeline was the same as that in healthy subjects. MS patients' WM lesions were manually segmented by identifying hyper-intensity regions in FLAIR images. The segmented lesions were further transformed to the DWI space by using FLIRT and FNIRT \citep{jenkinson2001flirt,andersson2007fnirt}. The NAWM was segmented in MPRAGE images by using FAST \citep{zhang2001fast} and transformed into the DWI space. To avoid partial volume effect, we excluded voxels close to MS lesions and CSF by expanding the mask of lesions and CSF by one voxel. An example of ROIs of MS lesions and NAWM is shown in \figref{fig:ms}a.

\subsection*{Statistics and Reproducibility}
The normality of distributions of inner axonal diameters in cross-sections with or without the presence of mitochondria was tested by using Anderson-Darling test, with a null hypothesis of normal distribution at 0.05 significance level; the null hypothesis was rejected for both diameter distributions with P-values $<$ 0.001. Further, the two diameter distributions were compared using one-sided Wilcoxon sum-rank test, with the null hypothesis that axonal diameters with the presence of mitochondria are not larger than those without. The significance level is 0.05. 

Eigenvalues and axial diffusivity were calculated voxel by voxel and averaged over each ROI. To evaluate the strength of axial diffusivity time-dependence in healthy subjects, we assumed that $D(t)$ is a linear function of $1/\sqrt{t}$ based on \eqref{eq:Dt-overall}, and calculated P-values with the null hypothesis of no positive correlation (one-sided test). Both the time-dependent parameters ($D_\infty$, $c$) in WM lesions and NAWM of MS patients did not pass the normality test (Anderson-Darling test) and were therefore compared using a paired one-sided Wilcoxon signed-rank test. For bulk diffusivity $D_\infty$, the null hypothesis is that $D_\infty$ in lesions is not larger than that in NAWM; for strength $c$ of restrictions, the null hypothesis is that $c$ in lesions is not smaller than that in NAWM. The significance level is 0.05.

In this study, we chose 1-tailed non-parametric test as we specifically hypothesized a decrease in the strength $c$ of restrictions in MS lesions as compared with NAMW, and an increase in the bulk diffusivity $D_\infty$. Indeed, since (1) the inner axonal diameters in cross-sections with the presence of mitochondria are larger than those without based on the previous histological study \citep{wang2003varicosity} and \figref{fig:text-mito}c, and (2) mitochondria density is increased in MS lesions due to demyelination \citep{witte2014msmito}, more variation in axon caliber is expected with a corresponding decrease in $c$. Similarly, we expect that $D_\infty$ in MS lesions is larger than that in NAWM due to demyelination \citep{moll2011ms}.
We would like to note that the MS data is rather exploratory due to the small sample size ($n$ = 5), which increases the risk of type 2 errors. In addition, the smallest possible P-value of a one-sided Wilcoxon signed rank test with $n$ = 5 is 0.03125, which provides a lower-bound for the P-value in this study.

\section*{Acknowledgements}
We thank 
Thorsten Feiweier for developing advanced diffusion WIP sequence, and Bigpurple High Performance Computing Center of New York University Langone Health for numerical computations on the cluster. Research was supported by the National Institute of Neurological Disorders and Stroke of the NIH under award number R01 NS088040 and R21 NS081230 and by the National Institute of Biomedical Imaging and Bioengineering (NIBIB) of the NIH under award number U01EB026996, and was performed at the Center of Advanced Imaging Innovation and Research (CAI2R, www.cai2r.net), an NIBIB Biomedical Technology Resource Center (NIH P41 EB017183).

\figref{fig:ias-shape}a-b is adapted by permission from Springer Nature: Springer Nature, Brain Structure and Function. Along-axon diameter variation and axonal orientation dispersion revealed with 3D electron microscopy: implications for quantifying brain white matter microstructure with histology and diffusion MRI. Hong-Hsi Lee, Katarina Yaros, Jelle Veraart, Jasmine L Pathan, Feng-Xia Liang, Sungheon G Kim, Dmitry S Novikov, and Els Fieremans. Copyright, 2019.\citep{lee2019em}

\section*{Data availability}
The SEM data and axon segmentation can be downloaded on our web page (http://cai2r.net/resources/software). All human brain MRI data for this study are available upon request.

\section*{Code availability}
The source codes of Monte Carlo simulations can be downloaded on our github page (https://github.com/NYU-DiffusionMRI).

\section*{Author contributions}
H.H.L., D.S.N. and E.F. designed research; 
H.H.L., D.S.N. and E.F. designed simulations; 
H.H.L. performed simulations; 
H.H.L. and D.S.N. developed theory; 
D.S.N and E.F. designed experiments; 
H.H.L., E.F. and A.P. performed experiments; 
H.H.L. and S.L.K. performed segmentations; 
H.H.L. analyzed data; 
D.S.N. and E.F. supervised the project;
D.S.N., E.F. and H.H.L. wrote the paper.

\section*{Competing interests}
The authors declare no competing interests.

\end{document}


\title{A time-dependent diffusion MRI signature of axon caliber variations and beading\\Supplementary Information}

\author{Hong-Hsi Lee}
\email{Honghsi.Lee@nyulangone.org}
\author{Antonios Papaioannou}
\author{Sung-Lyoung Kim}
\author{Dmitry S. Novikov}
\author{Els Fieremans}
\address{Center for Biomedical Imaging and Center for Advanced Imaging Innovation and Research (CAI$^2$R), Department of Radiology, New York University School of Medicine, New York, NY 10016, USA}

\begin{abstract}

\end{abstract}

\date{\today}


\maketitle

\beginsupplement

\subsection*{\normalsize Supplementary Notes}

\noindent{\bf Statistics of mitochondrial morphology.} For segmented mitochondria in Fig. 6a, the mitochondrial surface area is 2.26 $\pm$ 2.11 $\mu$m$^2$, and the mitochondrial volume is 0.21 $\pm$ 0.25 $\mu$m$^3$. For an individual axon, the number of mitochondria per unit IAS volume is 0.32 $\pm$ 0.14 $\mu$m$^{-3}$, the ratio of mitochondrial surface area to IAS volume is 0.67 $\pm$ 0.32 $\mu$m$^{-1}$, and the volume fraction of mitochondria to IAS is 6.0 $\pm$ 3.0\%, with histograms shown in Supplementary \figref{fig:app-mito}. All of these values are consistent with previous histological study in mouse optic nerve.\citep{stahon2016mitochondria}

Consequently, although the small mitochondrial volume ($\sim$6\% of the IAS volume) suggests a relatively small effect on the dMRI signal, as shown in Fig. 2b-c, they indirectly may alter the diffusion time-dependence, since the mitochondrial distribution along axons correlates with axon caliber variation. 

\bigskip\bigskip
\noindent{\bf Human brain data of 10 additional subjects.} The dMRI measurement was performed on 10 healthy subjects (7 males/3 females, 23-30 years old) by using a monopolar PGSE sequence provided by the vendor (Siemens WIP 511E) on a 3T Siemens Prisma scanner (Erlangen Germany) with a 64-channel head coil. For each subject, we varied diffusion time $t$ = [21.2, 22, 24, 26, 28, 30, 40, 50, 75, 100] ms and fixed diffusion gradient pulse width $\delta$ at 15 ms. For each scan, we obtained one $b$ = 0 non-diffusion weighted image and 64 DWIs of b-values $b$ = [0.1, 0.4, 1, 1.5] ms/$\mu$m$^2$ along [4, 10, 20, 30] gradient directions for each b-shell, with an isotropic resolution (2 mm)$^3$ and a field-of-view (216 mm)$^2$. The scanned brain volume was a slab of 15 slices, aligned parallel to the anterior commissure to posterior commissure line. The CC was in the middle of the slab for covering the entire CC. All scans were performed with the same TR/TE = 5000/150 ms. Total acquisition time is $\sim$65 min for each subject.

Image processing pipeline and chosen ROIs are the same as the ones in the main text.

The time-dependent axial diffusivity $D(t)$, measured by monopolar PGSE in the human brain WM (Supplementary \figref{fig:2mm-AD}a-b), were averaged over 10 healthy subjects and plotted with respect to $1/\sqrt{t}$. In all WM ROIs except Midbody of CC, the axial diffusivity time-dependence demonstrates a $1/\sqrt{t}$ power-law relation in equation (1) (P-value $<$ 0.05, Supplementary \tabref{tab:brain-pgse-2mm}), indicating that the universality class along WM axons is the short-range disorder (randomly distributed tissue inhomogeneity) in 1{\it d}, corresponding to a dynamical exponent $\vartheta$ = 1/2. The fitted parameters ($c$, $D_\infty$) are shown in Supplementary \tabref{tab:brain-pgse-2mm}.

Furthermore, the axial kurtosis in WM is $\sim$0.8, demonstrating the non-Gaussian diffusion along axons (Supplementary \figref{fig:2mm-AD}c-d).

\bigskip\bigskip
\noindent{\bf The effect of axonal diameter distribution on axial diffusivity time-dependence.} To investigate the effect of axonal diameter distribution on diffusivity time-dependence along axons, we performed Monte Carlo simulations in three different fiber bundles, composed of (1) realistic axonal shapes with caliber variations and axonal undulations (geometry I in Fig. 2a), (2) synthetic fibers with only axonal undulations (geometry IV in Fig. 2a), and (3) perfectly straight cylinders. The three fiber bundles have the same 2{\it d} cross-sectional diameter distribution and orientation dispersion based on Watson distribution. Simulation results in Supplementary \figref{fig:ias-und-cyl} demonstrate that fiber bundles with no caliber variations along individual fibers have very small diffusivity time-dependence along the bundles at clinical diffusion time $t=20-100$ ms, even for highly dispersed case ($\sim0.5$\% diffusivity change for $\theta=45^\circ$). In contrast, realistic axonal shapes with caliber variations along each axon show significant diffusivity $1/\sqrt{t}$-dependence along axons ($\sim5$\% diffusivity change for $\theta=45^\circ$ at $t=20-100$ ms). To sum up, the diffusivity time-dependence along axons is mainly contributed by caliber variations along individual axons, instead of axonal diameter distribution across different axons.

\cleardoublepage

\subsection*{\normalsize Supplementary Tables}

\begin{table}[bh!]
\small 
\centering
\begin{tabular}{l|ccc}
\hline
ROI      & P-value & $D_\infty$ ($\mu$m$^2$/ms) & $c$ ($\mu$m$^2\cdot$ms$^{-1/2}$) \\ \hline
ACR      & 6.4e-3  & 1.258 (0.019)             & 0.329 (0.105)                   \\
SCR      & 2.8e-4  & 1.328 (0.013)             & 0.413 (0.071)                   \\
PCR      & 1.3e-3  & 1.439 (0.013)             & 0.299 (0.074)                   \\
PLIC     & 4.0e-7  & 1.539 (0.006)             & 0.442 (0.033)                   \\
Genu     & 1.2e-2  & 1.484 (0.026)             & 0.389 (0.148)                   \\
Midbody  & 0.21    & -                & -                       \\
Splenium & 8.9e-3  & 1.777 (0.018)             & 0.299 (0.103)                   \\
ALIC     & 1.2e-2  & 1.400 (0.017)             & 0.298 (0.101)                   \\ \hline
\end{tabular}
\caption{\textnormal{Fit parameters of the time-dependent axial diffusivity $D(t)$ in human brain data measured using monopolar PGSE (Supplementary \figref{fig:2mm-AD}a-b). Standard errors are shown in the parenthesis. (ACR/SCR/PCR = anterior/superior/posterior corona radiate, ALIC/PLIC = anterior/posterior limb of the internal capsule, genu/midbody/splenium of CC)}}
\label{tab:brain-pgse-2mm}
\end{table}

\subsection*{\normalsize Supplementary Figures}

\begin{figure*}[th!!]\centering
\includegraphics[width=0.75\textwidth]{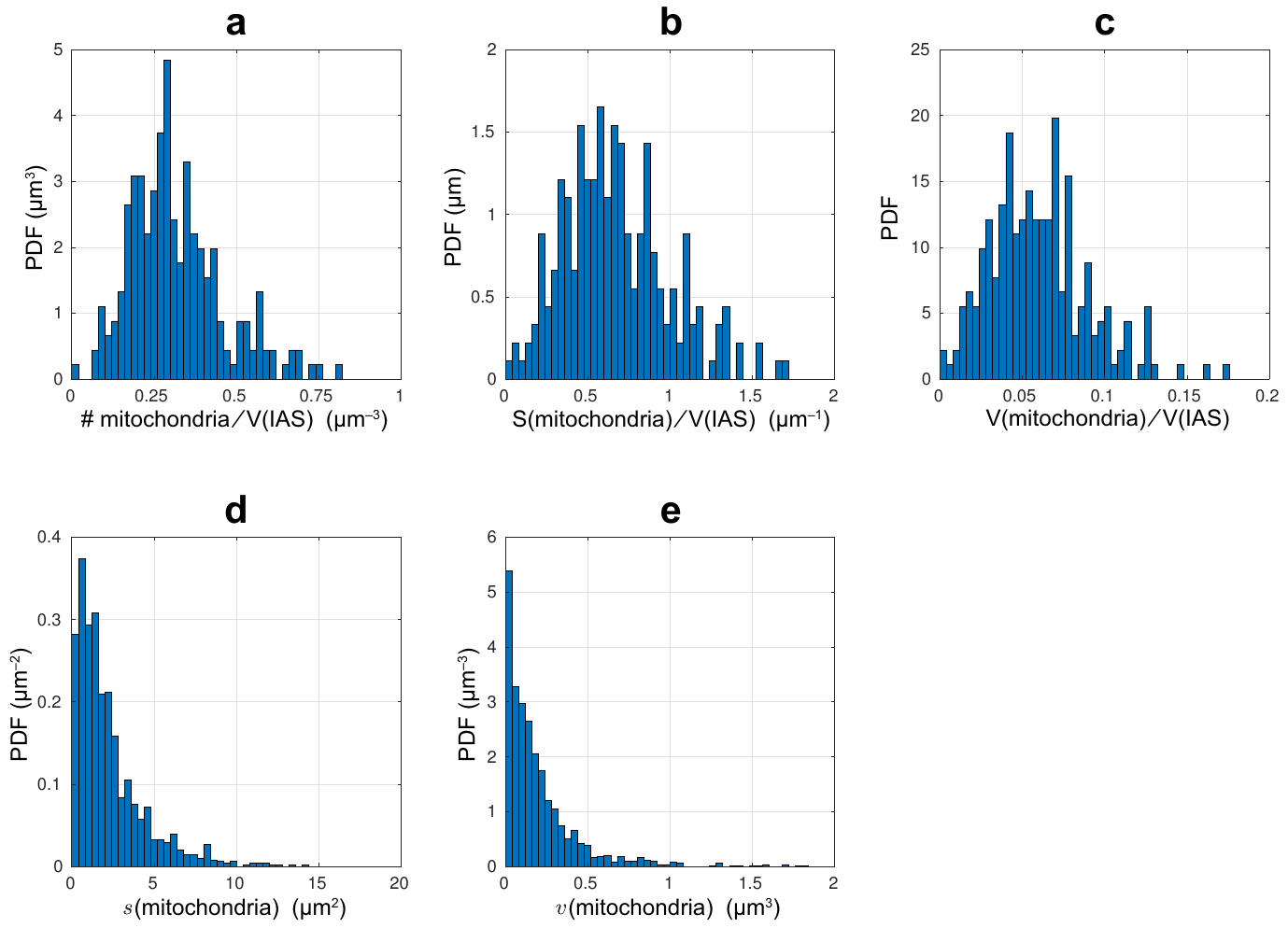}
\caption{Mitochondrial morphometry based on the segmentation in Fig. 6a:
\textnormal{{\bf a} Histogram of the mitochondrial number per unit IAS volume for each axon. {\bf b} Histogram of the ratio of mitochondrial surface area to IAS volume for each axon. {\bf c} Histogram of the volume fraction of mitochondria to IAS for each axon. {\bf d} Histogram of mitochondrial surface area of all segmented mitochondria. {\bf e} Histogram of mitochondrial volume of all segmented mitochondria.}}
\label{fig:app-mito}
\end{figure*}

\begin{figure*}[tb!]\centering
\includegraphics[width=0.77\textwidth]{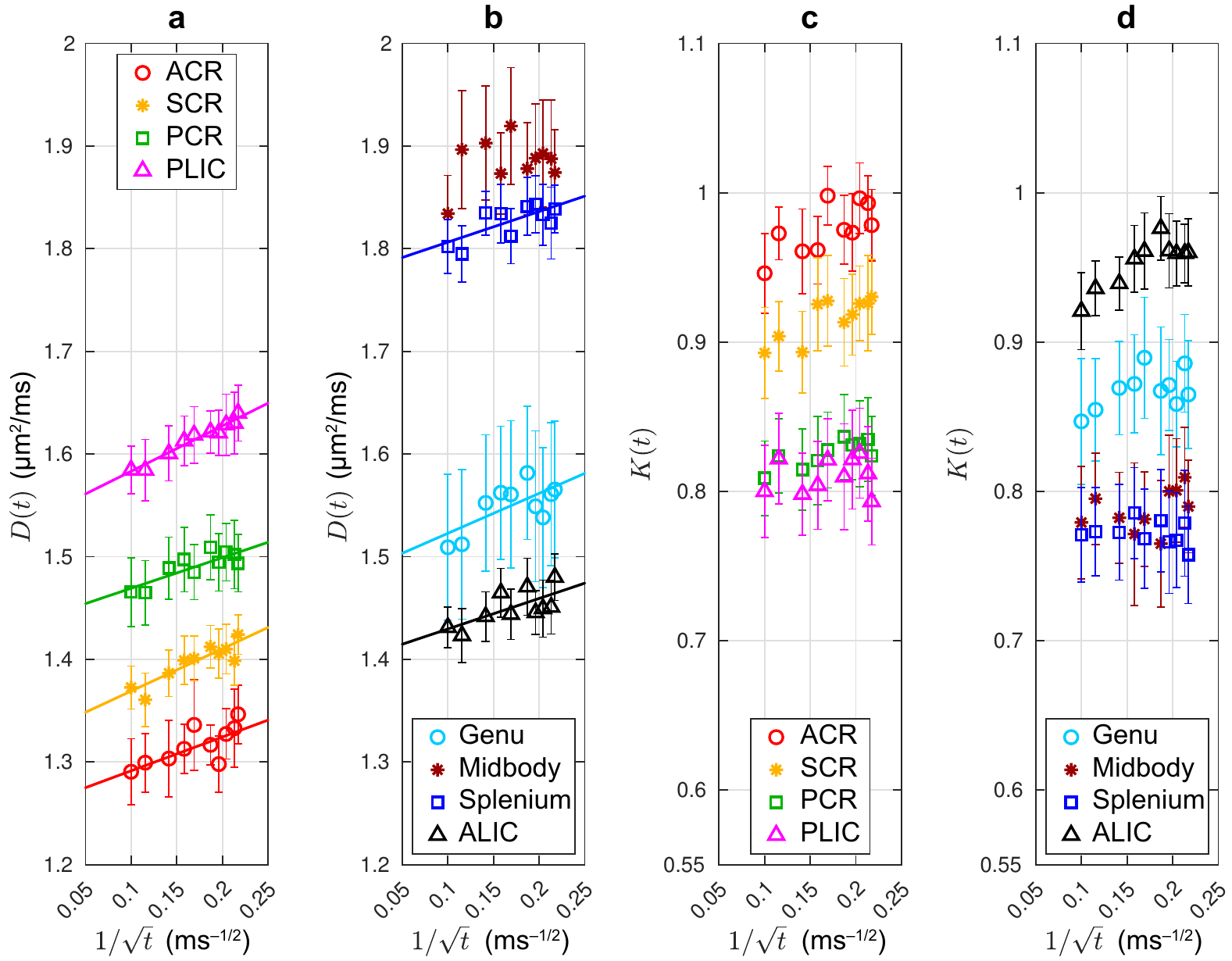}
\caption{\textnormal{{\bf a}, {\bf b} Time-dependent axial diffusivity $D(t)$ measured in vivo in brain WM of 10 healthy subjects using monopolar PGSE. In all WM ROIs except Midbody of CC, the experimental axial diffusivity scales as $1/\sqrt{t}$ (P-value $<$ 0.05, Supplementary \tabref{tab:brain-pgse-2mm}), manifesting that the universality class along WM axons is short-range disorder in 1{\it d}, corresponding to a power-law tail with $\vartheta$ = 1/2. The fit parameters are summarized in Supplementary \tabref{tab:brain-pgse-2mm}. {\bf c}, {\bf d} Time-dependent axial kurtosis $K(t)$ measured in vivo in brain WM of 10 healthy subjects using monopolar PGSE. The in vivo measured $K(t)$ is not zero, signifying the non-Gaussian diffusion along WM axons in the human brain.  The error bar indicates the standard error of 10 subjects. (ACR/SCR/PCR = anterior/superior/posterior corona radiate, ALIC/PLIC = anterior/posterior limb of the internal capsule, genu/midbody/splenium of CC)}}
\label{fig:2mm-AD}
\end{figure*}

\begin{figure*}[bt!]
\centering
	\includegraphics[width=0.96\textwidth]{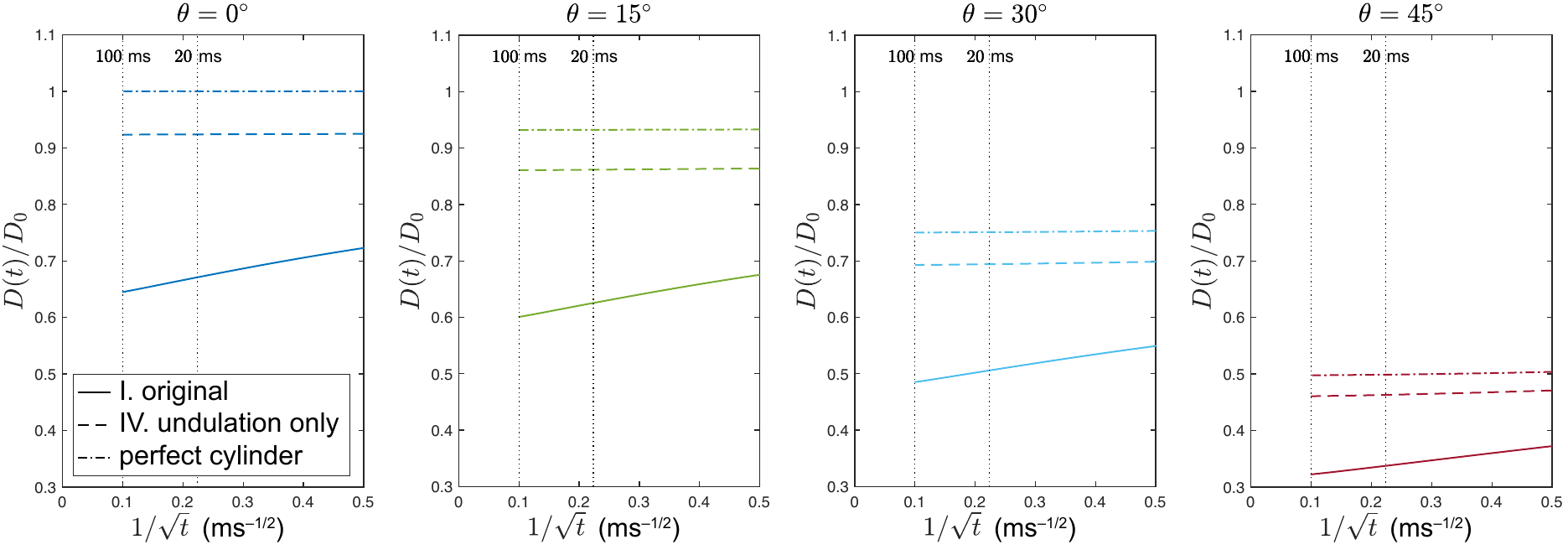}
	\caption[]{\textnormal{To demonstrate that the diffusivity time-dependence along axons is mainly contributed by caliber variations along individual axons, rather than diameter distribution across different axons, we performed Monte Carlo simulations in fiber bundles composed of (1) realistic IAS with caliber variations and axonal undulations (geometry I in Fig. 2a), (2) fibers with only axonal undulations (geometry IV in Fig. 2a), and (3) perfectly straight cylinders. The above three fiber bundles have the same 2{\it d} cross-sectional diameter distribution and orientation dispersion (Watson distribution). Simulation results show that fiber bundles with no caliber variations along individual fibers have negligible diffusivity time-dependence along the bundles at clinical diffusion time $t = 20-100$ ms, whereas realistic IAS has significant diffusivity $1/\sqrt{t}$-dependence along axons within the same time range.}}
	\label{fig:ias-und-cyl}
\end{figure*}